\documentclass[12pt,a4paper]{iopart}
\usepackage{iopams,setstack,mathrsfs,amsthm,cite,float,graphicx}
\newcommand{\al}{\alpha}
\newcommand{\be}{\beta}
\newcommand{\de}{\delta}

\newcommand{\vep}{\varepsilon}
\newcommand{\ga}{\gamma}
\newcommand{\ka}{\kappa}
\newcommand{\la}{\lambda}
\newcommand{\om}{\omega}
\newcommand{\si}{\sigma}

\newcommand{\vp}{\varphi}

\newcommand{\ze}{\zeta}
\newcommand{\bsi}{{\boldsymbol{\si}}}

\newcommand{\CC}{{\mathbb C}}
\newcommand{\NN}{{\mathbb N}}

\newcommand{\ZZ}{{\mathbb Z}}
\newcommand{\cE}{{\mathcal E}}

\newcommand{\cH}{{\mathcal H}}

\newcommand{\cN}{{\mathcal N}}

\newcommand{\cZ}{{\mathcal Z}}
\newcommand{\sS}{\mathscr S}
\newcommand{\ket}[1]{|#1\rangle}

\let\ds\displaystyle

\newcommand{\mss}{\kern 1pt}

\renewcommand{\le}{\leqslant}
\renewcommand{\ge}{\geqslant}
\newcommand{\tends}[1]{\bbuildrel{\hbox to 2em{\rightarrowfill}}_{#1}^{}}
\newcommand{\operatorname}[1]{\mathop{\rm #1}\nolimits}
\newcommand{\sech}{\operatorname{sech}}
\newcommand{\erf}{\operatorname{erf}}

\let\iu\rmi
\let\diff\rmd

\newcommand{\su}{\mathrm{su}}

\renewcommand{\Im}{\operatorname{Im}}

\newcommand{\en}{\enspace}
\newcommand{\all}{\forall}

\theoremstyle{remark}
\newtheorem{rem}{Remark}%
\let\tfrac\case
\let\eqref\eref
\let\text\mbox
\newcommand{\multiparteqlabel}[1]{\addtocounter{equation}{1}
  \eqnarray\nonumber
  \endeqnarray\label{#1}\addtocounter{equation}{-1}\vspace*{-42pt}}
\eqnobysec
\begin{document}
\title[Exact solution and thermodynamics of a spin chain with long-range interactions]{Exact
  solution and thermodynamics of a spin chain with long-range elliptic interactions}
\author{Federico Finkel and Artemio Gonz\'alez-L\'opez} \address{Departamento de F\'\i sica Te\'orica II,
  Universidad Complutense de Madrid, 28040 Madrid, Spain} \eads{\mailto{ffinkel@ucm.es} and
  \mailto{artemio@ucm.es}}
\date{\today}
\begin{abstract}
  We solve in closed form the simplest ($\su(1|1)$) supersymmetric version of Inozemtsev's
  elliptic spin chain, as well as its infinite (hyperbolic) counterpart. The solution relies on
  the equivalence of these models to a system of free spinless fermions, and on the exact
  computation of the Fourier transform of the resulting elliptic hopping amplitude. We also
  compute the thermodynamic functions of the finite (elliptic) chain and their low temperature
  limit, and show that the energy levels become normally distributed in the thermodynamic limit.
  Our results indicate that at low temperatures the $\su(1|1)$ elliptic chain behaves as a
  critical $X\!X$ model, and deviates in an essential way from the Haldane--Shastry chain.
\end{abstract} {\it Keywords\/}: integrable spin chains (vertex models), solvable lattice models
\maketitle

\setcounter{footnote}{0}
%
%
\section{Introduction}
The spin chain introduced independently by Haldane and Shastry in 1988 is one of the most widely
studied integrable lattice models with long-ranged interactions~\cite{Ha88,Sh88}. This chain
describes $N$ equidistant spins on a circle, with two-body interactions inversely proportional to
the square of the distance measured along the chord. The Hamiltonian of the Haldane--Shastry (HS)
chain can be written as
\begin{equation}\label{HS}
  H=\frac{J\pi^2}{N^2}\,\sum\limits_{i<j}\sin^{-2}\biggl(\frac\pi N(i-j)\biggr)\,(1-S_{ij}),
\end{equation}
where $S_{ij}$ is the operator permuting the $i$-th and $j$-th spins, and the indices in the
double sum range from $1$ to $N$. If the Hilbert space of each spin is $m$-dimensional, the
operator $S_{ij}$ can be easily expressed in terms of the $\su(m)$ spin operators of the $i$-th
and $j$-th spins~\cite{Po94}. In particular, for spin $1/2$ ($m=2$) we have
\[
S_{ij}=\frac12\,(1+\bsi_i\cdot\bsi_j)\,,
\]
where $\bsi_k\equiv(\si_k^x,\si_k^y,\si_k^z)$ denotes the three Pauli matrices acting on the
$k$-th spin. In fact, there is also a natural $\su(m|n)$ supersymmetric version of the HS chain,
which has been extensively studied in the literature~\cite{Ha93,GW95,BB06}.

The HS chain possesses remarkable physical and mathematical properties. Indeed, it is intimately
related to the one-dimensional Hubbard model with long-range hopping, from which it is obtained in
the limit of infinite on-site interaction at half filling~\cite{GR92}. It has also been proposed
as the simplest model providing an explicit realization of fractional statistics and anyons in one
dimension~\cite{Ha91b,GS05,Gr09,HHTBP92}. As to its mathematical properties, the HS chain is
completely integrable~\cite{FM93}, can be solved via an asymptotic Bethe
ansatz~\cite{Ha91,Ka92,HH93}, is one of the few models invariant under the Yangian group for a
finite number of sites~\cite{HHTBP92}, and its partition function can be evaluated in closed
form~\cite{FG05} via Polychronakos's ``freezing trick''~\cite{Po93}.

In fact, the HS chain~\eqref{HS} is essentially a limiting case of a more general model due to
Inozemtsev~\cite{In90}, whose Hamiltonian we shall take as
\begin{equation}\label{In}
  H=J\sum_{i<j}\wp_N(i-j)(1-S_{ij})\,.
\end{equation}
Here
\[
\wp_N(x)\equiv\wp\biggl(x;\frac N2,\frac{\iu\al}2\biggr)
\]
denotes the Weierstrass elliptic function~\cite{WW27} with periods $N$ and $\iu\al$, with $\al>0$.
Indeed, since
\begin{equation}\label{toin}
  \lim_{\al\to\infty}\wp_N(x)
  =\frac{\pi^2}{N^2}\bigg(\sin^{-2}\biggl(\frac{\pi x}N\biggr)-\frac13\bigg)
\end{equation}
(see~\ref{app.limits}), the $\al\to\infty$ limit of the Hamiltonian~\eqref{In} differs
from~\eqref{HS} by a constant multiple of the operator $\sum_{i<j}(1-S_{ij})$, which trivially
commutes with the HS chain Hamiltonian. More surprising is the fact that when $\al\to0$ the
Inozemtsev chain is related to the Heisenberg chain
\begin{equation}\label{He}
  H=J\sum_{i=1}^N(1-S_{i,i+1})\,,\qquad S_{N,N+1}\equiv S_{1N}
\end{equation}
(cf.~\cite{In90}). More precisely , if $ 1\le x\le N-\nobreak1$ we have
\begin{equation}\label{to0}
  \lim_{\al\to0}\e^{2\pi/\al}\bigg(\frac{\al^2}{4\pi^2}\,\wp_N(x)-\frac1{12}\bigg)=\de_{1,x}+
  \de_{N-1,x}
\end{equation}
(see again~\ref{app.limits}).

Interestingly, in the last decade the Inozemtsev chain has also received considerable attention in
the context of the AdS/CFT correspondence~\cite{Ma98,Wi98,GKP98,BMN02}. This unexpected connection
stems from the work of Minahan and Zarembo~\cite{MZ03}, who showed that at one loop the (planar)
spectrum of the dilation operator of $\cN=4$ super Yang--Mills gauge theory can be generated by a
suitable integrable spin chain with nearest-neighbor interactions. In order to generalize the
latter result to more than one loop, it becomes necessary to consider integrable spin chains with
long-range interactions. Moreover, any such chain describing $\cN=4$
Berenstein--Maldacena--Nastase theory non-perturbatively should contain an additional parameter
related to the Yang--Mills coupling constant. As first noted by Serban and Staudacher~\cite{SS04},
the Inozemtsev chain~\eqref{In} is one of the simplest models fulfilling the last two
requirements. Since this chain is also generally believed to be integrable, it has been
extensively studied as a candidate for generating the planar spectrum of the dilation operator at
several loops~\cite{Be12}. It should be noted, however, that a complete rigorous proof of the
integrability of the Inozemtsev chain has not yet been found, in spite of several promising
partial results in this direction~\cite{In90,In03}. In any case, the energy spectrum of this chain
is not explicitly known beyond the two-magnon sector.

In this paper we introduce an $\su(1|1)$ elliptic chain, which is in fact the simplest
supersymmetric version of the Inozemtsev model~\eqref{In}. We show that, rather unexpectedly, this
chain is completely integrable and its whole spectrum can be computed in closed form. Our proof is
based on two key ideas. In the first place, we exploit the well-known fact that the $\su(1|1)$
permutation operators can be expressed in terms of annihilation and creation operators of a single
species of spinless fermions. This implies that the $\su(1|1)$ elliptic chain is equivalent to a
model of hopping free fermions, which smoothly interpolates between the $\su(1|1)$ HS chain and
the standard $X\!X$ model (at a critical value of the magnetic field). The second ingredient in
our proof is the explicit computation of the dispersion relation of the $\su(1|1)$ elliptic chain
using standard techniques in analytic function theory~\cite{In90,In03,FG14}.

From the explicit knowledge of the dispersion relation, we have been able to study several
properties of the $\su(1|1)$ elliptic chain in the thermodynamic limit. In the first place, we
have analyzed the low momentum behavior of the dispersion relation, showing that the energy is
quadratic in the momentum near $p=0$ for all finite values of the chain's parameter $\al$. Thus,
at low energies the $\su(1|1)$ elliptic chain behaves as a critical $X\!X$ model, and in
particular its low energy excitations cannot be described by an effective two-dimensional
conformal field theory. By contrast, it is well known that the dispersion relation of the
$\su(1|1)$ HS chain is linear near the origin, and at low energies its spectrum coincides with
that of a conformal field theory of $m$ noninteracting Dirac fermions with only positive
energies~\cite{BBS08}. We have next computed the chain's thermodynamic functions in closed form,
and determined their low temperature limit. Our analysis shows that at low temperatures the
$\su(1|1)$ elliptic chain behaves essentially as a critical $X\!X$ model, and markedly differs
from both the $\su(1|1)$ and $\su(2)$ HS chains. We have also studied the asymptotic behavior of
the level density as the number of sites tends to infinity, proving that it approaches a Gaussian
distribution with parameters equal to the mean and the standard deviation of the spectrum, as is
typically the case for spin chains of HS type~\cite{EFG09,EFG10,BB06,BB09}. Finally, we have
introduced the $\su(1|1)$ supersymmetric analog of Inozemtsev's infinite (hyperbolic) chain,
showing that its dispersion relation is proportional to the thermodynamic limit of the $\su(1|1)$
elliptic chain's dispersion relation.

The paper is organized as follows. In Section~\ref{sec.su11chain} we introduce the $\su(1|1)$
elliptic chain and, as explained above, solve it by transforming it into a system of hopping
fermions whose dispersion relation we compute in closed form. Section~\ref{sec.thermo} is devoted
to the derivation of the chain's thermodynamic functions and the analysis of the asymptotic
behavior of its level density. Section~\ref{sec.infchain} deals with the $\su(1|1)$ version of
Inozemtsev's infinite hyperbolic chain and its complete solution. In Section~\ref{sec.concout} we
summarize the paper's main results, and point out several future developments suggested by our
work. The paper ends with two technical appendices on the computation of several limits involving
elliptic Weierstrass functions, and on the proof of the properties of quasi-periodic functions
needed for the evaluation of the dispersion relation of the $\su(1|1)$ elliptic chain.

\section{The $\su(1|1)$ elliptic chain and its solution}\label{sec.su11chain}
The model we shall study is the $\su(1|1)$ version of Inozemtsev's chain~\eqref{In}, in which each
site is occupied by a spinless particle which can be either a boson or a fermion. Thus the chain's
Hamiltonian can be written as
\multiparteqlabel{FGh}
\numparts
\begin{equation}\label{FG}
  H=J\sum\limits_{i<j}h(i-j)(1-\sS_{ij}),
\end{equation}
where, as before,
\begin{equation}
  h(x)=\wp_N(x)\,,
  \label{hwPN}
\end{equation}
\endnumparts
and the $\sS_{ij}$ are the standard $\su(1|1)$ permutation
operators. More precisely, if we denote by $\ket 0$ and $\ket 1$ respectively the boson and the
fermion states, the standard basis of the chain's Hilbert space consists of the $2^N$ product
states
\[
\ket{s_1}\otimes\cdots\otimes\ket{s_N}\equiv\ket{s_1,\dots,s_N}\,,\qquad s_i\in\{0,1\}\,.
\]
We then have
\[
\sS_{ij}\ket{s_1,\dots,s_i,\dots,s_j,\dots,s_N}=(-1)^n\ket{s_1,\dots,s_j,\dots,s_i,\dots,s_N}\,,
\]
where $n=s_i=s_j$ if $s_i=s_j$, while $n$ is equal to the number of fermions occupying the sites
$i+1,\dots,j-1$ if $s_i\ne s_j$. Equivalently,
\[
\sS_{ij}=b^\dagger_ib^\dagger_j b_ib_j+f^\dagger_if^\dagger_j f_if_j+b^\dagger_if^\dagger_j
f_ib_j+f^\dagger_ib^\dagger_j b_if_j\,,
\]
where $b_k^\dagger$ and $f_k^\dagger$ respectively denote the boson and fermion creation operators
acting on the $k$-th site. It is important to note that, since each site is occupied by either one
boson or one fermion, the chain's Hilbert space is the subspace $\cH$ of the Fock space determined
by the constraints
\begin{equation}\label{constraints}
  b^\dagger_ib_i+f^\dagger_if_i=1\,,\qquad i=1,\dots,N\,.
\end{equation}

We shall next show that a chain of the form~\eqref{FG} with \emph{arbitrary} $h$ can be mapped to
a system of $N$ ``hopping'' (spinless) free fermions. The key idea is to regard the bosonic state
$\ket0$ as the vacuum for the fermion. More formally, we define the operators
\[
a_i^\dagger\equiv f^\dagger_ib_i\,,\qquad i=1,\dots,N\,,
\]
and note that they satisfy the canonical anticommutation relations on $\cH$ on account of the
constraints~\eqref{constraints}. Indeed, it is immediate to check that
$\big\{a_i^\dagger,a^\dagger_j\big\}=\big\{a_i,a_j\big\}=0$, while
\[
\big\{a_i^\dagger,a_j\big\}=\big\{f_i^\dagger,f_j\big\}b_j^\dagger b_i +f_i^\dagger
f_j\big[b_i,b_j^\dagger\big]=\de_{ij}\big(b_i^\dagger b_i +f^\dagger_if_i\big)=\de_{ij}
\quad\text{on }\cH\,.
\]
Thus $a_i^\dagger$ creates a fermion at the site $i$. The chain's sites can now be
either empty or occupied by a fermion, and the Hilbert space $\cH$ is identified with the
\emph{whole} Fock space for the new system of fermions. Our next task is to express the $\su(1|1)$
permutation operator $\sS_{ij}$ in terms of the fermionic operators $a_k,a_k^\dagger$. To this
end, note first of all that
\[
b^\dagger_if^\dagger_j f_ib_j+f^\dagger_ib^\dagger_j b_if_j=a^\dagger_ja_i+a^\dagger_ia_j\,.
\]
On the other hand, using the constraints~\eqref{constraints} we easily obtain
\[
a^\dagger_ia_i=f^\dagger_if_ib_ib^\dagger_i=f^\dagger_if_i(1+b_i^\dagger
b_i)=2f^\dagger_if_i-(f^\dagger_if_i)^2 =f^\dagger_if_i\quad\text{on }\cH\,,
\]
so that
\begin{eqnarray*}
  b^\dagger_ib^\dagger_j b_ib_j+f^\dagger_if^\dagger_j
  f_if_j&=(1-f_i^\dagger f_i)(1-f_j^\dagger f_j)-f^\dagger_if_if^\dagger_jf_j=1-f_i^\dagger
          f_i-f_j^\dagger f_j\\&=1-a^\dagger_ia_i-a^\dagger_ja_j\quad\text{on }\cH\,.
\end{eqnarray*}
We thus have
\begin{equation}\label{Sij}
  \sS_{ij}=1-a^\dagger_ia_i-a^\dagger_ja_j+a^\dagger_ia_j+a^\dagger_ja_i\,,
\end{equation}
as first noted by Haldane~\cite{Ha93}. Substituting~\eqref{Sij} into~\eqref{FG} we easily obtain
\begin{equation}
  H=J\sum_{i\ne j}h(|i-j|)a^\dagger_i(a_i-a_j)\,,
\label{Hnotrinv}
\end{equation}
which is indeed the Hamiltonian of a system of $N$ free hopping fermions.

The Weierstrass function $h(x)=\wp_N(x)$ is even and $N$-periodic, i.e,
\begin{equation}
  h(x)=h(-x)=h(x+N)\,,\qquad\all x\,.
  \label{hcond}
\end{equation}
Hence
\begin{equation}
  \label{trinv}
  h(x)=h(N-x)\,,\qquad\all x\,,
\end{equation}
so that the model~\eqref{FGh} is translation-invariant. Equivalently, the chain sites can be
viewed as $N$ equidistant points lying on a circle, as in the case of the Haldane--Shastry chain.
When the chain~\eqref{FG} is translation-invariant, i.e., when the function $h$
satisfies~\eqref{trinv}, the diagonal terms in Eq.~\eqref{Hnotrinv} can be considerably
simplified. Indeed, in this case the coefficient of $a^\dagger_ia_i$ is given by
\begin{equation}\label{sum21}\fl
\sum_{j;j\ne i}h(|i-j|)=\sum_{l=1}^{i-1}h(l)+\sum_{l=1}^{N-i}h(l)
=\sum_{l=1}^{i-1}h(l)+\sum_{l=i}^{N-1}h(N-l)=\sum_{l=1}^{N-1}h(l)\,,
\end{equation}
independently of $i$, so that we can write
\begin{equation}
  \label{FGf}
  H=-J\sum_{i,j=1}^Nh(|i-j|)a^\dagger_i a_j\,,
\end{equation}
provided that we set
\begin{equation}\label{h0}
h(0)\equiv-\sum_{l=1}^{N-1}h(l)\,.
\end{equation}
Thus a translation-invariant chain~\eqref{FG}-\eqref{trinv} is equivalent to a system of $N$ free
hopping fermions lying on a circle, with hopping amplitude between the $i$-th and
$j$-th sites equal to $h(|i-j|)$.

When $h$ satisfies Eq.~\eqref{trinv}, the Hamiltonian~\eqref{FGf} can be diagonalized by
performing the discrete Fourier transform
\begin{equation}\label{FT}
  c_l=\frac1{\sqrt N}\,\sum_{k=1}^N\e^{-2\pi\iu kl/N}a_k\,,\qquad l=0,1,\dots,N-1\,.
\end{equation}
Indeed, first of all it is immediate to check that the operators $c_l$ satisfy the canonical
anticommutation relations, on account of the unitarity of the mapping~\eqref{FT}; in fact, we
shall show below that $c_l^\dagger$ creates a fermion with momentum $2\pi l/N$ ($\bmod\en 2\pi$).
Using the inverse Fourier transform formula
\[
a_k=\frac1{\sqrt N}\sum_{l=0}^{N-1}\e^{2\pi\iu kl/N}c_l\,,\qquad
k=1,\dots,N\,,
\]
in the Hamiltonian~\eqref{FGf} we obtain
\[
H = J\sum_{l,m=0}^{N-1}h_{lm}c_l^\dagger c_m\,,
\]
with
\begin{eqnarray*}\fl
  h_{lm}&=-\frac1N\sum_{j,k=1}^Nh(|k-j|)\,\e^{2\pi\iu(jm-kl)/N}\\\fl
        &=-\frac1N\bigg(\sum_{s=0}^{N-1}\e^{-2\pi\iu
          sl/N}h(s)\sum_{j=1}^{N-s}\e^{2\pi\iu
          j(m-l)/N}+\sum_{s=1}^{N-1}\e^{2\pi\iu
          sl/N}h(s)\sum_{j=s+1}^N\e^{2\pi\iu
          j(m-l)/N}
          \bigg)\,.
\end{eqnarray*}
Performing the change of index $s\mapsto N-s$ in the last term and using Eq.~\eqref{trinv} we
easily obtain
\[
h_{lm}=-\frac1N\sum_{s=0}^{N-1}\e^{-2\pi\iu
  sl/N}h(s)\sum_{j=1}^{N}\e^{2\pi\iu j(m-l)/N}
=\de_{lm}\vep_l\,,
\]
with
\[
\vep_l=-\sum_{s=0}^{N-1}\e^{-2\pi\iu sl/N}h(s)\,.
\]
Thus
\begin{equation}\label{Hep}
  H=J\sum_{l=0}^{N-1}\vep_l\,c_l^\dagger c_l\,,
\end{equation}
is indeed diagonal when written in terms of the Fourier-transformed operators $c_l$. The
dispersion relation~$\vep_l$ can be further simplified using the definition of $h(0)$ and
Eq.~\eqref{trinv}, namely
\begin{eqnarray}\label{vepl}
  \vep_l&=-h(0)-\sum_{j=1}^{N-1}\e^{-2\pi\iu jl/N}h(j)=\sum_{j=1}^{N-1}\Big(1-\e^{-2\pi\iu
          jl/N}\Big)h(j)\nonumber\\&=\sum_{j=1}^{N-1}\Big(1-\e^{2\pi\iu
          jl/N}\Big)h(j)
          =
\sum_{j=1}^{N-1}\Big(1-\cos\bigl(2\pi jl/N\bigr)\Big)\,h(j)\,.
\end{eqnarray}
\begin{rem}
  From the latter equation it immediately follows that
  \[
  \vep_0=0\,,\qquad \vep_l=\vep_{N-l}\,.
  \]
  By the first of these identities and Eq.~\eqref{Hep}, the degeneracy of each energy level is
  even, as required by the $\su(1|1)$ symmetry.
\end{rem}
\begin{rem}
  Another immediate consequence of Eq.~\eqref{Hep} is the complete integrability of the
  model~\eqref{FG}-\eqref{trinv}, since the number operators $c^\dagger_lc_l$ $(l=0,\dots,N-1$)
  are a commuting family of first integrals.
\end{rem}

\begin{rem}
  Consider the translation operator $T$, defined on the basis states by
  \[
  T\ket{s_1,\dots,s_N}=\ket{s_2,\dots,s_N,s_1}\,.
  \]
  It is easy to check that $T$ is characterized by the relations
  \begin{equation}
    T^{-1}a_jT=a_{j+1}\,,\quad j=1,\dots,N\,;\qquad T\ket0=\ket0\,,\label{Tconds}
  \end{equation}
  where $\ket0\equiv\ket{0,\dots,0}$ denotes the vacuum state. The momentum operator $P$ is
  defined (up to integer multiples of $2\pi$) in the usual way:
  \[
  T=\e^{\iu P}\,.
  \]
  It is immediate to check that when the function $h$ satisfies Eq.~\eqref{trinv} the
  Hamiltonian~\eqref{FGf} commutes with $T$, and hence with $P$. We shall next show that $P$ is
  also diagonalized by the Fourier transform. Indeed, note first of all that
  conditions~\eqref{Tconds} are equivalent to
  \[
  T^{-1}c_l\,T=\e^{2\pi\iu\,l/N}c_l\,,\qquad l=0,\dots,N-1\,.
  \]
  If we make the ansatz
  \[
  P=\sum_{l=0}^{N-1}p_l\,c^\dagger_lc_l\,,
  \]
  the latter conditions obviously reduce to
  \[
  c_l\exp\big(\iu\,p_l\,c^\dagger_lc_l\big)=
  \e^{2\pi\iu\,l/N}\exp\big(\iu\,p_l\,c^\dagger_lc_l\big)\,c_l\,,
  \qquad l=0,\dots,N-1\,.
  \]
  The RHS is clearly equal to
  \[
  \e^{2\pi\iu\,l/N}c_l\,,
  \]
  since $c_l^2=0$. On the other hand, from the identity
  \[
  c_l(c^\dagger_lc_l)^k=c_l\,,\qquad k\in\NN\,,
  \]
  it follows that the LHS equals
  \[
  \sum_{k=0}^\infty\frac{(\iu p_l)^k}{k!}\,c_l=\e^{\iu p_l}c_l\,.
  \]
  Hence
  \[
  p_l=\frac{2\pi l}N\quad ({\bmod}\en2\pi)\,,
  \]
  and the momentum operator $P$ is explicitly given by
  \[
  P=\frac{2\pi}N\,\sum_{l=0}^{N-1}l\,c^\dagger_lc_l\quad ({\bmod}\en2\pi)
  \]
  (cf.~Ref.~\cite{BBS08}). Thus the state created by $c_l^\dagger$ has well-defined energy
  $\vep_l$ and momentum $2\pi l/N$ (${\bmod}\en 2\pi$), as we had anticipated.
\end{rem}

\smallskip As we have just seen, in order to solve the elliptic chain~\eqref{FGh} we need to
evaluate in closed form the sum in Eq.~\eqref{vepl} when~$h(x)=\wp_N(x)$.
In fact, the sum $\sum_{j=1}^N\wp_N(j)$ was computed in Ref.~\cite{FG14}, with the result
\begin{equation}
  \label{wpNsum}
  \sum_{j=1}^{N-1}\wp_N(j)=\frac2{\iu\al}\,\bigg[\eta_3\biggl(\frac12,\frac{\iu\al}2\biggr)
  -N\eta_3\biggl(\frac N2,\frac{\iu\al}2\biggr)\bigg]\,.
\end{equation}
Here we have used the standard notation
\[
\eta_i(\om_1,\om_3)\equiv\ze(\om_i;\om_1,\om_3)\,,
\]
where $\ze(z;\om_1,\om_3)$ denotes the Weierstrass zeta function associated to the lattice
$2m\om_1+2n\om_3$, with $m,n\in\ZZ$ and $\Im(\om_3/\om_1)>0$ (cf.~\cite{WW27}).
It thus suffices to compute the discrete Fourier transform of the Weierstrass function $\wp_N(x)$.
This can be done using a technique due to Inozemtsev~\cite{In90,In03}, which we shall briefly
summarize for the reader's convenience.

For fixed $l=1,2,\dots,N-1$, we define the function
\[
f_l(z)=\sum_{j=0}^{N-1}\e^{-2\pi\iu jl/N}\,\wp_N(z+j)\,.
\]
From the periodicity of the Weierstrass elliptic function, it is immediate to check that $f_l$
satisfies the quasi-periodicity conditions in Eq.~\eqref{fqp} with
\[
2\om_1=1\,,\qquad2\om_3=\iu\al\,,\qquad p=l/N\,.
\]
On the other hand, $f_l$ is clearly analytic everywhere except at points congruent to the origin,
i.e., on the lattice $m+\iu n\al$ ($m,n\in\ZZ$). In fact, the only term in the sum defining $f_l$
which is singular at the origin is the one with $j=0$. Since\footnote{We shall write
  $f(z)=\Or\bigl(g(z)\bigr)$ for $z\to z_0$ if there is a positive constant $C$ such that
  $|f(z)|\le C|g(z)|$ for $z$ sufficiently close to $z_0$.}
\begin{equation}\label{wpL}
  \wp(z;\om_1,\om_3)=\frac1{z^2}+\Or(z^2)\,,
\end{equation}
the Laurent series of $f_l$ about $z=0$ is simply
\[
f_l(z)=\frac1{z^2}+\Or(1)\,,
\]
and we can therefore apply Eqs.~\eqref{fz} and~\eqref{z0term} with $\om_3=\iu\al/2$ and $p=l/N$.
We thus have
\begin{equation}\label{limf1z2}
  \lim_{z\to0}\bigg(f_l(z)-\frac1{z^2}\bigg)=\frac12\,\wp_1\biggl(\frac{\iu\al l}N\biggr)-
  \frac12\,\bigg(\ze_1\biggl(\frac{\iu\al l}N\biggr)-\frac
  {2l}N\,\ze_1\biggl(\frac{\iu\al}2\biggr)\bigg)^2\,,
\end{equation}
where $\wp_1$ and $\ze_1$ are the Weierstrass functions with half-periods $1/2$ and $\iu\al/2$.
On the other hand, from the definition of
$f_l$ and Eq.~\eqref{wpL} we easily obtain
\[
f_l(z)=\frac1{z^2}+\sum_{j=1}^{N-1}\e^{-2\pi\iu jl/N}\,\wp_N(j)+\Or(z)\,.
\]
Comparing with Eq.~\eqref{limf1z2} we conclude that
\[
  \sum_{j=1}^{N-1}\e^{-2\pi\iu jl/N}\,\wp_N(j)=\frac12\,\wp_1\biggl(\frac{\iu\al l}N\biggr)-
  \frac12\,\bigg(\ze_1\biggl(\frac{\iu\al l}N\biggr)-\frac
  {2l}N\,\ze_1\biggl(\frac{\iu\al}2\biggr)\bigg)^2\,.
\]
Combining this formula with Eq.~\eqref{wpNsum} we obtain the following explicit expression for the
dispersion relation $\vep_l$ for $l=1,\dots,N-1$:
\begin{equation}\label{eplim}\fl
\vep_l=\frac2{\iu\al}\,\bigg[\ze_1\biggl(\frac{\iu\al}2\biggr) -N\eta_3\biggl(\frac
N2,\frac{\iu\al}2\biggr)\bigg]-\frac12\,\bigg[\wp_1\biggl(\frac{\iu\al l}N\biggr)-
\bigg(\ze_1\biggl(\frac{\iu\al l}N\biggr)-\frac
{2l}N\,\ze_1\biggl(\frac{\iu\al}2\biggr)\bigg)^2\bigg]
\end{equation}
(and, of course, $\vep_0=0$). This formula can be further simplified with
the help of the homogeneity properties of the Weierstrass functions, namely
\begin{equation}\label{homog}\fl
\wp(\la z;\la\om_1,\la\om_3)=\frac1{\la^2}\,\wp(z;\om_1,\om_3)\,,\qquad
\ze(\la z;\la\om_1,\la\om_3)=\frac1{\la}\,\ze(z;\om_1,\om_3)\,.
\end{equation}
Using these formulas with $\la=\iu\al$ we easily arrive at the following closed-form expression for the dispersion relation of the
$\su(1|1)$ elliptic chain~\eqref{FGh}:
\begin{equation}\fl\label{eplraw}
  \vep_l=\frac1{2\al^2}\,\bigg[\wp(l/N)-\bigg(\ze(l/N)
  -2\eta_1\frac
  lN\bigg)^2+4\big(N\hat\eta_1-\eta_1\big)\bigg]\,,\qquad l=1,\dots,N-1\,,
\end{equation}
where from now on
\[
\wp(z)\equiv\wp\biggl(z;\frac12,\frac\iu{2\al}\biggr)\,,\qquad
\ze(z)\equiv\ze\biggl(z;\frac12,\frac\iu{2\al}\biggr)
\]
shall denote the Weierstrass functions with periods $1$ and $\iu/\al$, and we have also set
\[
\qquad\eta_1\equiv\eta_1\biggl(\frac12,\frac{\iu}{2\al}\biggr)\,,\qquad
\hat\eta_1\equiv\eta_1\biggl(\frac12,\frac{\iu N}{2\al}\biggr)\,.
\]

\begin{rem}
  The reader may have noticed that Eq.~\eqref{eplim} coincides with the dispersion relation of the
  $1$-magnons with momentum $2\pi l/N$ ($\bmod\ 2\pi$) of the $\su(2)$ elliptic chain~\eqref{In}
  computed in Ref.~\cite{In90}. In fact, this remarkable relation between the $\su(2)$ and
  $\su(1|1)$ elliptic chains holds for an arbitrary translation-invariant chain of the
  form~\eqref{FG}. This is a further indication of the greater simplicity of the $\su(1|1)$ models
  compared to their $\su(2)$ counterparts.
\end{rem}

\begin{figure}[h]
  \centering
  \includegraphics[width=.6\columnwidth]{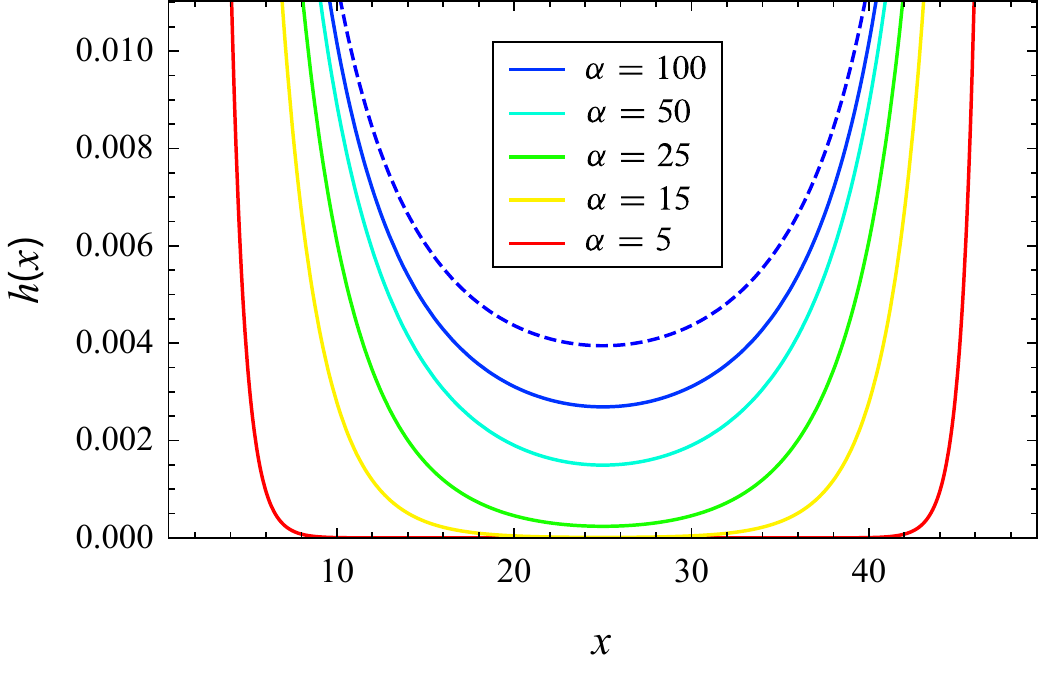}
  \caption{(Color online) Interaction strength $h(x)$ in Eq.~\eqref{newh} for $N=50$ spins and
    several values of $\al$ in the range $[5,100]$. The dashed blue line represents the
    interaction strength $h(x)=(\pi/50)^2\sin^{-2}(\pi x/50)$ of the Haldane--Shastry chain with
    $50$ spins ($\al=\infty$).}
  \label{fig.h}
\end{figure}

\begin{figure}
  \centering
  \includegraphics[width=.6\columnwidth]{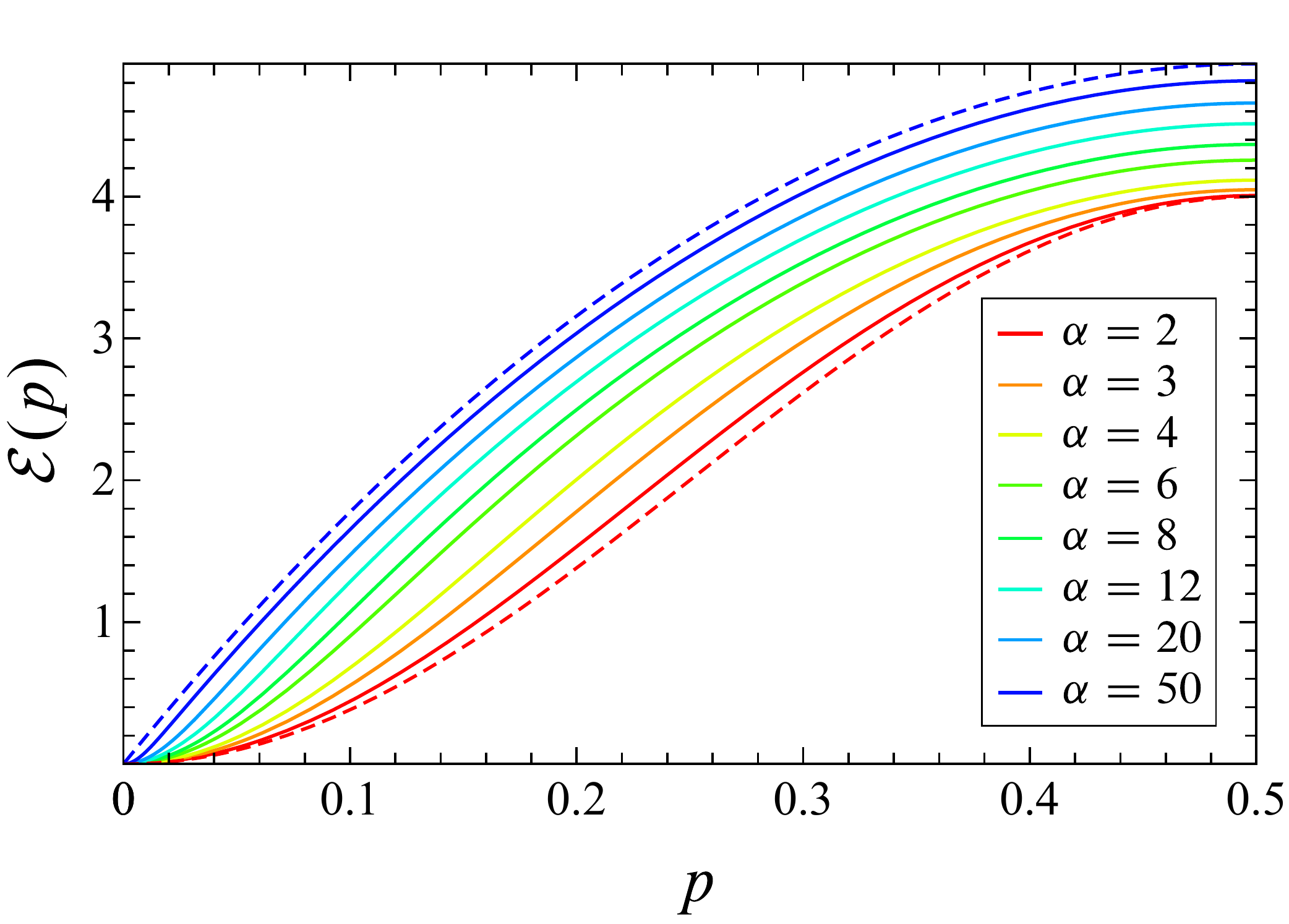}
  \caption{(Color online). Solid lines: dispersion relation~\eqref{cE} of the elliptic $\su(1|1)$
    chain~\eqref{FG}-\eqref{newh} for several values of the parameter~$\al$ between $2$ (bottom)
    and $50$ (top). Dashed lines: dispersion relations of the the critical $X\!X$ model~\eqref{XX}
    (bottom) and the $\su(1|1)$ Haldane--Shastry chain (top). Only the range $0\le p\le 1/2$ has
    been shown, since $\cE(p)=\cE(1-p)$ on account of~\eqref{vepl}.}
  \label{fig.cEp}
\end{figure}

All the terms in the dispersion relation~\eqref{eplraw} depend on $N$ through the combination
$l/N$ except for the one involving $\hat\eta_1$, which can be eliminated through the shift
$h(x)\mapsto h(x)-(2\hat\eta_1/\al^2)$. In fact, it is possible to combine this shift with an
appropriate rescaling so that the $\al\to\infty$ and $\al\to0$ limits of the chain~\eqref{FG}
exactly coincide with the $\su(1|1)$ versions of the Haldane--Shastry and Heisenberg chains, i.e.,
Eqs.~\eqref{HS} and~\eqref{He} with $S_{ij}$ replaced by $\sS_{ij}$. Indeed, it suffices to
take~\cite{In05}
\begin{equation}\label{newh}
  h(x)=\bigg(\frac\al\pi\bigg)^2\sinh^2\biggl(\frac\pi\al\biggr)\,\bigg(\wp_N(x)
  -\frac{2\hat\eta_1}{\al^2}\bigg)\,;
\end{equation}
see~\ref{app.limits} for the details. For this reason, we shall from now on adopt the
normalization~\eqref{newh} when referring to the $\su(1|1)$ elliptic chain~\eqref{FG}. By
Eq.~\eqref{vepl}, the corresponding dispersion relation is simply\footnote{It can be easily
  checked (cf.~Eq.~\eqref{mal} below) that $\cE(0)\equiv\lim_{p\to0}\cE(p)=0$.}
\multiparteqlabel{disprel}\numparts
\begin{equation}\label{vepcep}
\vep_l=\cE(l/N)\,,
\end{equation}
where
\begin{equation}\label{cE}
  \cE(p)=\frac{\sinh^2(\pi/\al)}{2\pi^2}\,\Big[\wp(p)-\big(\ze(p)-2\eta_1 p\big)^2-4\eta_1\Big]
\end{equation}
\endnumparts
is independent of $N$, and $p$ is the physical moment in units of $2\pi$.
\begin{rem}
  It should be noted that the rescaled interaction strength $h(x)$ in Eq.~\eqref{newh} remains
  positive for $0<x<N$, so that the chain~\eqref{FG}-\eqref{newh} is of ferromagnetic type for
  $J>0$ (cf.~Fig.~\ref{fig.h}). Indeed, the function $\wp_N(x)$ has an absolute minimum in the interval $0<x<N$ at the
  real half-period $x=N/2$, and thus it suffices to show that
  \[
  \wp_N(N/2)-\frac{2\hat\eta_1}{\al^2}>0\,.
  \]
  From Eq.~\eqref{wpseries} with $\om_1=\iu\al/2$, $\om_3=-N/2$ (so that
  $\Im(\om_3/\om_1)=N/\al>0$) and the homogeneity of the Weierstrass zeta function (cf.~the second
  Eq.~\eqref{homog}) we easily obtain
  \begin{eqnarray*}\fl
    \wp_N(N/2)-\frac{2\hat\eta_1}{\al^2}&=\wp_N(N/2)+\frac{2}{\iu\al}\,\eta_1(\iu\al/2,-N/2)\equiv
    \wp(\om_3;\om_1,\om_3)+\frac1{\om_1}\eta_1(\om_1,\om_3)\\
    &=\frac{\pi^2}{\al^2}\sinh^{-2}\biggl(\frac{N\pi}{2\al}\biggr)
      +\frac{4\pi^2}{\al^2}\sum_{n=1}^\infty n\,\e^{-nN\pi/\al}\,
      \coth\biggl(\frac{nN\pi}{\al}\biggr)>0\,.
  \end{eqnarray*}
\end{rem}
It should be expected (and can, in fact, be analytically proved; see~\ref{app.limits} for the
details) that the $\al\to\infty$ and $\al\to0$ limits of Eq.~\eqref{cE} respectively yield the
dispersion relations of the $\su(1|1)$ Haldane--Shastry and Heisenberg chains. In fact, the
dispersion relation of the $\su(1|1)$ HS chain was computed in Ref.~\cite{GW95} essentially by the
same procedure followed here, with the result
\begin{equation}\label{dispHS}
\cE(p)=2\pi^2 p(1-p)\,.
\end{equation}
As to the $\su(1|1)$ Heisenberg chain, note first of all that from Eq.~\eqref{FGf} with
$h(x)=\de_{1,x}+\de_{N-1,x}$ (and thus $h(0)=-2$, by Eq.~\eqref{h0}) it immediately follows that
the Hamiltonian of this model can be expressed as
\[
H=J\Big[2\sum_{i=1}^Na^\dagger_ia_i-\sum_{i=1}^N(a^\dagger_ia_{i+1}+a^\dagger_{i+1}a_i)\Big]\,,\qquad
a_{N+1}\equiv a_1\,.
\]
It is well known~\cite{Sa11} that the latter Hamiltonian is transformed by the standard
Jordan--Wigner transformation
\[
a_k=\si_1^z\cdots\si_{k-1}^z\cdot\frac12(\si_k^x-\iu\si_k^y)\,,\qquad k=1,\dots,N\,,
\]
into the $X\!X$ model (at a critical value of the magnetic field)
\begin{equation}\label{XX}
  H=J\bigg[\frac12\sum_{i=1}^N\big(\si_i^x\si_{i+1}^x+\si_i^y\si_{i+1}^y\big)
  +\sum_{i=1}^N(1+\si_i^z)\bigg]\,,\qquad \bsi_{N+1}\equiv\bsi_{1}\,.
\end{equation}
It follows from the previous discussion that the $\al\to0$ limit of the $\su(1|1)$ elliptic
chain~\eqref{FG}-\eqref{newh} is equivalent to a critical $X\!X$ model, a fact that is not obvious
at all \emph{a priori}. From Eq.~\eqref{vepl} with $h(x)=\de_{1,x}+\de_{N-1,x}$, it is immediate
to obtain the well-known dispersion relation of this model
\begin{equation}\label{dispXX}
\cE(p)=4\sin^2(\pi p)\,.
\end{equation}
As expected, the dispersion relation~\eqref{cE} varies smoothly between its
limits~\eqref{dispXX} and~\eqref{dispHS} as $\al$ ranges from $0$ to $\infty$; see,
e.g.,~Fig.~\ref{fig.cEp}.

Let us next briefly analyze the low momentum behavior of the dispersion relation~\eqref{cE}. To
this end, we recall the Laurent series
\[\fl
\wp(z)=\frac1{z^2}+\frac{g_2}{20}\,z^2+\frac{g_3}{28}\,z^4+\Or(z^6)\,,\qquad
\ze(z)=\frac1z-\frac{g_2}{60}\,z^3-\frac{g_3}{140}\,z^5+\Or(z^7)\,,
\]
where $g_i\equiv g_i\big(1/2,\iu/(2\al)\big)$ are the invariants of the Weierstrass function with
periods $(1,\iu/\al)$ (see, e.g.,~\cite{WW27}). Expanding Eq.~\eqref{cE} around $p=0$ with the
help of the previous formulas we readily obtain
\begin{equation}\label{mal}
  \cE(p)=\frac{p^2}{2m(\al)}+O(p^4)\,,
\end{equation}
where the effective mass $m(\al)$ is given by
\[
m(\al)=\frac{12\pi^2}{(g_2-48\,\eta_1^2)\sinh^2(\pi/\al)}\,.
\]
As $\cE(p)=\cE(1-p)$, a similar formula is valid around $p=1$, with $p$ replaced by $1-p$. In
particular, since the low-energy dispersion relation is not linear in the momentum, the low energy
excitations cannot be described by an effective two-dimensional conformal field theory. By
contrast, it is well known that the low energy excitations of the $\su(m|1)$-supersymmetric
Haldane--Shastry spin chain coincide with the spectrum of a conformal field theory of $m$
non-interacting Dirac fermions with only positive energies~\cite{BBS08}.

\section{Thermodynamics}\label{sec.thermo}
\subsection{Thermodynamic functions}
Using the dispersion relation~\eqref{cE}, it is straightforward to evaluate the free energy per
site
\[
f(T)=-\frac1\be\lim_{N\to\infty}\frac{\log\cZ_N}{N}\,,
\]
where $\cZ_N$ is the partition function for $N$ spins. Indeed, since the model in momentum space
is equivalent to a system of $N$ free fermions with energies $\vep_l=J\cE(l/N)$, the partition
function is given by
\begin{equation}\label{cZN}
\cZ_N =\prod_{l=0}^{N-1}\big(1+\e^{-J\be\cE(l/N)}\big)\,,
\end{equation}
and therefore
\[\fl
f(T)=-\frac1\be\,\lim_{N\to\infty}\sum_{l=0}^{N-1}\frac1N\log\Bigl(1+\e^{-J\be\cE(l/N)}\Bigr)
=-\frac1\be\int_0^1\log\Bigl(1+\e^{-J\be\cE(p)}\Bigr)\,\diff p\,.
\]
Using this explicit formula it is immediate to compute in closed form the remaining thermodynamic
functions, i.e., the energy, the specific heat, and the entropy (per site), respectively given by
\[
u = \frac{\partial}{\partial\be}\,(\be f)\,,\qquad
c = \frac{\partial u}{\partial T}\,,\qquad
s = \frac1T\,(u-f)\,.
\]
Introducing the dimensionless temperature $\tau\equiv1/(|J|\,\be)$ we easily obtain:
\multiparteqlabel{TD} \numparts
\begin{eqnarray}
  \frac1{|J|}\,(f-u_0)=-\tau\int_0^1\log\big(1+\e^{-\cE(p)/\tau}\big)\,\diff p\label{f}\\
  \frac1{|J|}\,(u-u_0)=\int_0^1\frac{\cE(p)}{1+\e^{\cE(p)/\tau}}\,\diff p\label{u}\\
  \frac{c}{k_{\mathrm B}}=\frac1{4\tau^2}\int_0^1\cE^2(p)\sech^2\bigl(\tfrac{\cE(p)}{2\tau}\bigr)
  \,\diff
  p\\
  \frac{s}{k_{\mathrm B}}=\int_0^1\vp\bigl(\tfrac{\cE(p)}{2\tau}\bigr)\,\diff p\,,\label{s}
\end{eqnarray}
\endnumparts
where $\vp(x)\equiv\log(2\cosh x)-x\tanh x$ and
\[
u_0\equiv u(0)=f(0)=\frac 12\,(J-|J|)\int_0^1\cE(p)\,\diff p\,.
\]
Remarkably, the last integral can be evaluated in closed form, with the result
\begin{equation}\label{cEint}
\int_0^1\cE(p)=\frac2{\pi^2}\sinh^2(\pi/\al)\bigg(\frac{\pi^2}6-\eta_1\bigg)\,;
\end{equation}
see the next subsection for more details. Likewise, the density of fermions is given by
\[
\nu_{\mathrm F} = \lim_{N\to\infty}\frac1N\sum_{l=0}^{N-1}\big(1+\e^{J\be\cE(l/N)}\big)^{-1}=
\int_0^1\frac{\diff p}{1+\e^{J\be\cE(p)}}\,,
\]
or, in terms of the dimensionless temperature $\tau$,
\[
\frac{J}{|J|}\big(\nu_{\mathrm F}-\tfrac12\big)+\tfrac12=\int_0^1\frac{\diff
  p}{1+\e^{\cE(p)/\tau}}\,.
\]

\begin{figure}[h]
  \includegraphics[width=.475\columnwidth]{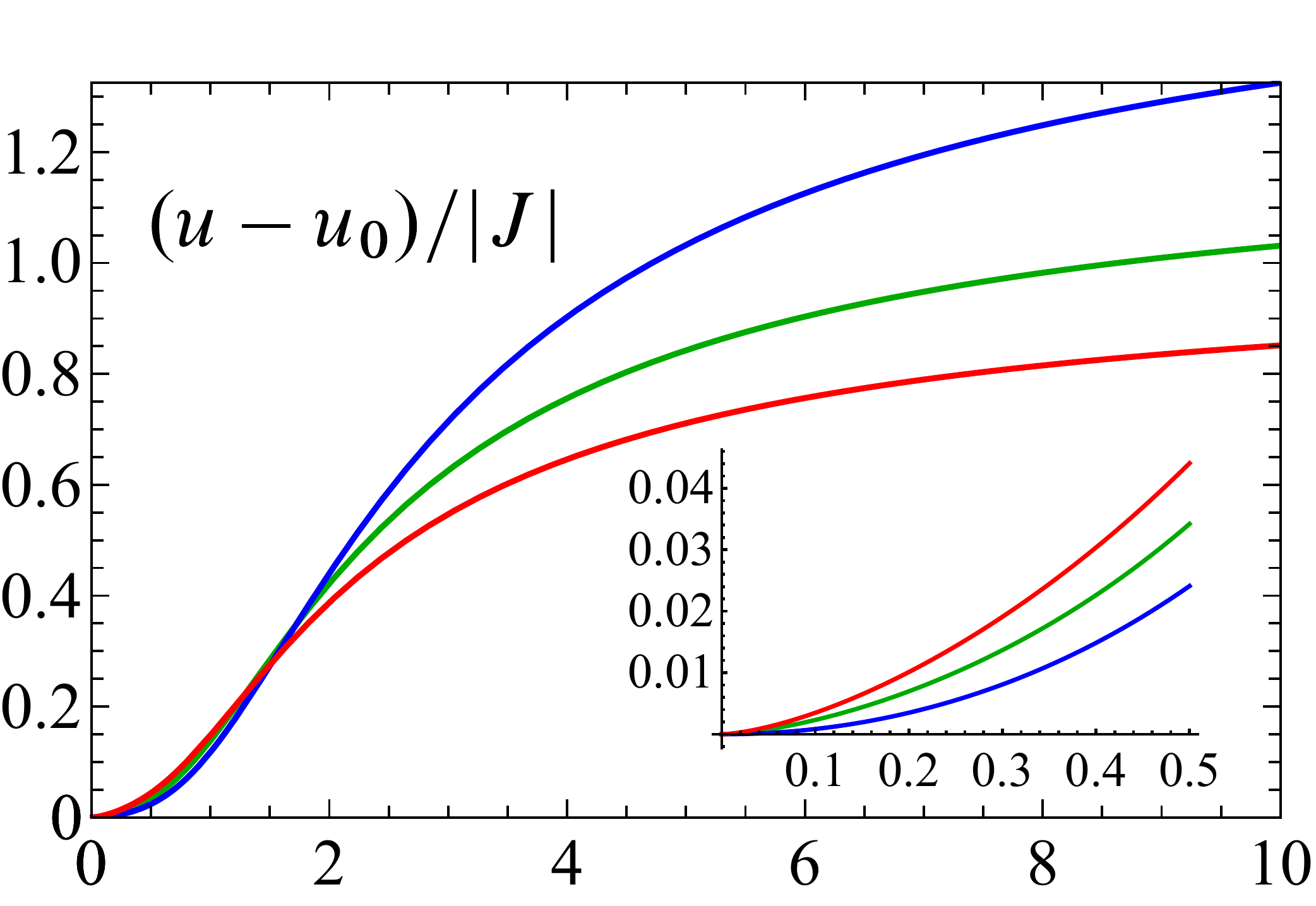}\hfill
  \includegraphics[width=.475\columnwidth]{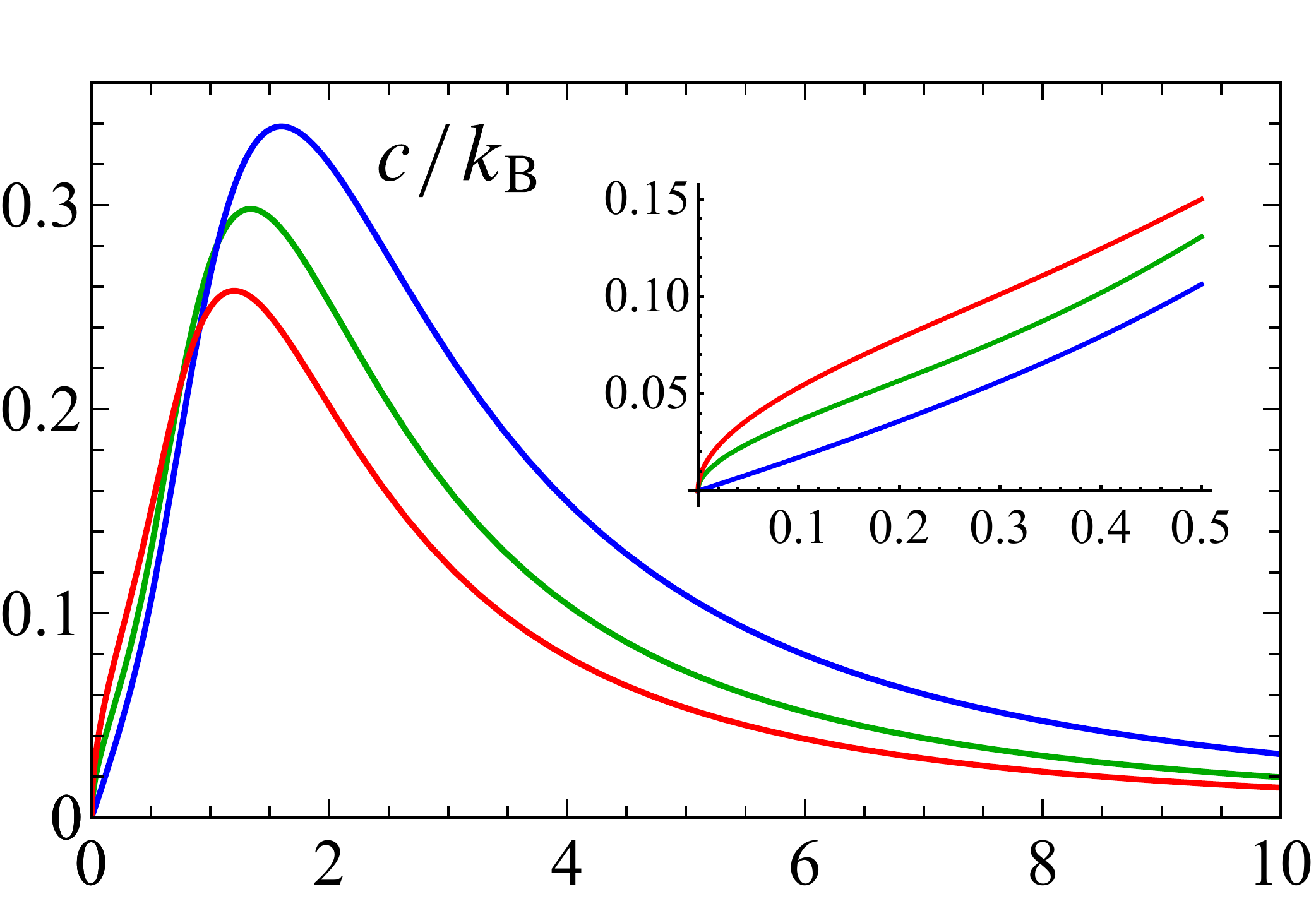}\\
  \includegraphics[width=.475\columnwidth]{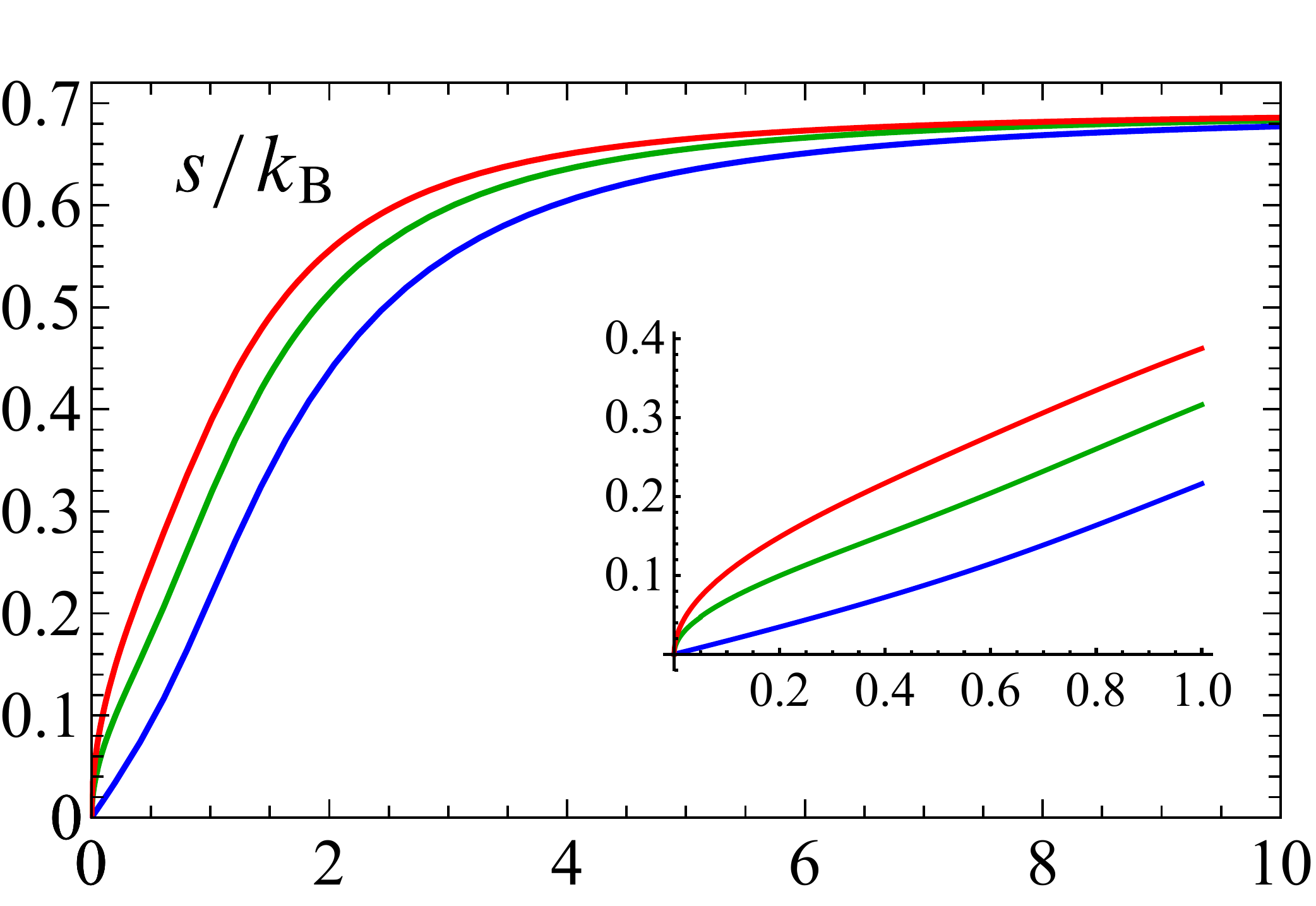}\hfill
  \includegraphics[width=.475\columnwidth]{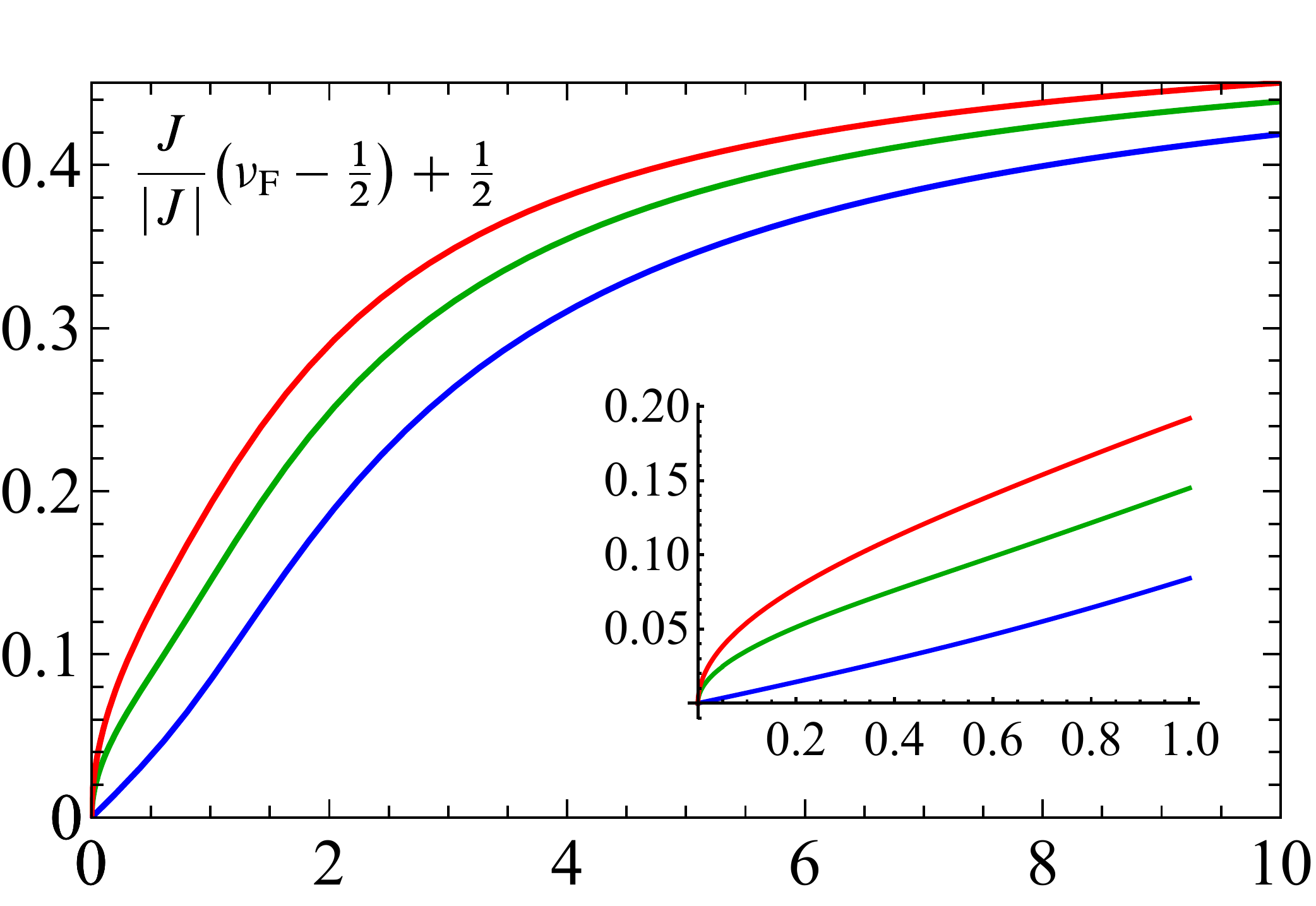}
   
  \caption{(Color online). Energy, specific heat, entropy (per site) and density of fermions for
    the elliptic $\su(1|1)$ chain with $\al=5$ (green), the critical $X\!X$ model~\eqref{XX}
    (red), and the $\su(1|1)$ Haldane--Shastry chain (blue) as a function of the dimensionless
    temperature $\tau=k_{\mathrm B}T/|J|$.}
  \label{fig.TD}
\end{figure}

The low temperature behavior of the thermodynamic functions can be readily deduced from the
previous formulas. Indeed, performing the change of variables $x=\cE(p)/\tau$ in Eq.~\eqref{f} we
obtain
\[\fl
\frac1{|J|}\,(f-u_0)=-2\tau\int_0^{1/2}\log\big(1+\e^{-\cE(p)/\tau}\big)\,\diff p=
-2\tau^2\int_0^{\cE(1/2)/\tau}\log(1+\e^{-x})\,\frac{\diff x}{\cE'(p)}\,,
\]
where we have taken into account the symmetry of $\cE$ under $p\mapsto 1-p$. From Eq.~\eqref{mal}
it follows that $p=\Or\bigl(\sqrt{\tau x}\,\bigr)$ and
\[
\cE'(p)=\frac p{m(\al)}+\Or(p^3)=\sqrt{\frac{2\mss\tau x}{m(\al)}}\,\big(1+\Or(\tau x)\big)\,.
\]
Hence
\begin{equation}
  \frac1{|J|}\,(f-u_0)=-\gamma\sqrt{m(\al)}\,\tau^{3/2}+\Or(\tau^{5/2})\,,
  \label{fasymp}
\end{equation}
where
\[
\gamma\equiv\sqrt2\int_0^\infty\frac{\log(1+\e^{-x})}{\sqrt x}\,\diff x
\]
is a numeric constant which can be readily computed in closed form. Indeed,
\begin{eqnarray*}
\ga
&=\sqrt2\,\sum_{k=1}^\infty\frac{(-1)^{k+1}}k\int_0^\infty x^{-1/2}\e^{-kx}\diff x
=\sqrt2\,\sum_{k=1}^\infty\frac{(-1)^{k+1}}{k^{3/2}}\int_0^\infty t^{-1/2}\e^{-t}\diff t\\
&=\sqrt{2\pi}\,\sum_{k=1}^\infty\frac{(-1)^{k+1}}{k^{3/2}}\equiv\sqrt{2\pi}\,\eta(3/2)\,,
\end{eqnarray*}
where $\eta(z)$ denotes the Dirichlet's eta function. Using the well-know
identity~(cf.~\cite{OLBC10}, Eq.~(25.2.3))
\[
\eta(z)=(1-2^{1-z})\ze_{\mathrm R}(z)\,,
\]
where $\ze_{\mathrm R}$ denotes Riemann's zeta function, we finally obtain
\[
\gamma =\big(\sqrt2-1\big)\sqrt\pi\,\ze_{\mathrm R}(3/2)\simeq1.91794\,.
\]
The asymptotic expansions of the remaining thermodynamic functions follow immediately from their
definition and Eq.~\eqref{fasymp}:
\begin{eqnarray*}
  \frac1{|J|}\,(u-u_0)=\frac \gamma2\sqrt{m(\al)}\,\tau^{3/2}+\Or(\tau^{5/2})\,,\\
  \frac{c}{k_{\mathrm B}}=\frac{3\ga}4\,\sqrt{m(\al)}\,\tau^{1/2}+\Or(\tau^{3/2})\,,\\
  \frac{s}{k_{\mathrm B}}=\frac{3\ga}2\,\sqrt{m(\al)}\,\tau^{1/2}+\Or(\tau^{3/2})\,.
\end{eqnarray*}
The low temperature behavior of the density of fermions can be computed in a similar way, with the
result
\[\fl
\frac{J}{|J|}\big(\nu_{\mathrm F}-\tfrac12\big)+\tfrac12
=\ga'\sqrt{m(\al)}\,\tau^{1/2}+\Or(\tau^{3/2})\,,
\qquad\ga'\equiv\sqrt{\pi}\,(\sqrt2-2)\,\ze_{\mathrm R}(\tfrac12)\simeq1.51626\,.
\]
Note that the previous formulas are valid for the limiting case $\al=0$, i.e., for the critical
$X\!X$ model~\eqref{XX}, whose effective mass is $m(0)=(8\pi^2)^{-1}$. Thus, at low temperatures
the $\su(1|1)$ elliptic chain~\eqref{FG}-\eqref{newh} is equivalent to the critical $X\!X$
model~\eqref{XX} rescaled by the factor $(8\pi^2m(\al))^{-1}$. Similarly, replacing $\cE(p)$ in
Eqs.~\eqref{TD} by its $\al\to\infty$ limit $2\pi^2p(1-p)$, we obtain the thermodynamic functions
of the $\su(1|1)$ Haldane--Shastry chain. The resulting formulas exactly coincide with those
deduced in Ref.~\cite{EFG12} for its $\su(2)$ counterpart. Thus, in the thermodynamic limit the
$\su(1|1)$ and $\su(2)$ Haldane--Shastry chains are equivalent (though this is certainly not the
case for any finite value of $N$). Note, however, that the low temperature behavior of these
models~\cite{EFG12} markedly differs from that of the elliptic $\su(1|1)$ chain
(cf.~Fig.~\ref{fig.TD}), since their dispersion relation is linear near $p=0$ and $p=1$.
\subsection{Level density}
One of the characteristic properties of the $\su(m)$ HS chain (and, indeed, of other related spin
chains with long-range interactions) is the fact that in the limit $N\to\infty$ its level density
(normalized to one) approaches a Gaussian distribution with parameters equal to the mean and the
standard deviation of the spectrum~\cite{EFG10}. By a suitable generalization of the central limit
theorem, it was shown in Ref.~\cite{EFG09} that this property also holds for systems with a
factorizable partition function, i.e., such that
\begin{equation}\label{cZNfact}
\cZ_N(T)=\prod_{l=0}^{N-1}\cZ_l(T;N)\,,
\end{equation}
provided only that two general conditions are satisfied. The first of these conditions simply
states that for sufficiently large $N$
\begin{equation}\label{sicond}
\si_l\le C N^{-1/2}\si\,,\qquad l=0,\dots,N-1\,,
\end{equation}
for some positive constant $C$ independent of $N$. In the latter formula $\si_l$ and $\si$
respectively denote the standard deviation of the spectrum of the $l$-th subsystem and of the
whole system, obviously related by
\[
\si^2=\sum_{l=0}^{N-1}\si_l^2\,.
\]
The second condition, which is of a rather more technical nature, is automatically satisfied when
each $\cZ_l$ is the partition function of a two-level system~\cite{EFG09}. By Eq.~\eqref{cZN},
this is exactly what happens for the $\su(1|1)$ elliptic chain~\eqref{FG}-\eqref{newh}, since in
this case~\eqref{cZNfact} holds with
\[
\cZ_l=1+\e^{-\be\cE(l/N)}\,.
\]
(For simplicity, in the latter formula and for the rest of this section we have set $J=1$.)
Moreover, from the latter equation we obviously have $\si_l=\frac12\,\cE(l/N)$, so that
condition~\eqref{sicond} reduces to
\begin{equation}\label{condfinal}
\frac1N\sum_{k=0}^{N-1}\cE(k/N)^2\ge C'\cE(l/N)^2\,,\qquad l=0,\dots,N-1\,,
\end{equation}
for some positive constant $C'$. These inequalities are clearly
satisfied for $N$ large enough. Indeed, on the one hand we have
\[
\cE(l/N)\le \cE_{\mathrm{max}}\,,\qquad l=0,\dots,N-1\,,
\]
where $\cE_{\mathrm{max}}$ is the maximum of $\cE(p)$ in the compact interval $[0,1]$, which is of
course independent of $N$. (It can be shown that $\cE_{\mathrm{max}}=\cE(1/2)$, although this
result is irrelevant for what follows.) On the other hand, since the LHS of Eq.~\eqref{condfinal}
tends to $\int_0^1\cE^2(p)\diff p$ as $N\to\infty$, there exists a natural number $N_0$ such that
\[
\frac1N\sum_{k=0}^{N-1}\cE(k/N)^2\ge\frac12\int_0^1\cE^2(p)\diff p\,,\qquad N\ge N_0\,.
\]
Thus condition~\eqref{condfinal} is satisfied for $N\ge N_0$ taking
\[
C'=\frac1{2\cE_{\mathrm{max}}^2}\,\int_0^1\cE^2(p)\diff p\,.
\]
\begin{figure}[h]
  \centering
  \includegraphics[width=.6\columnwidth]{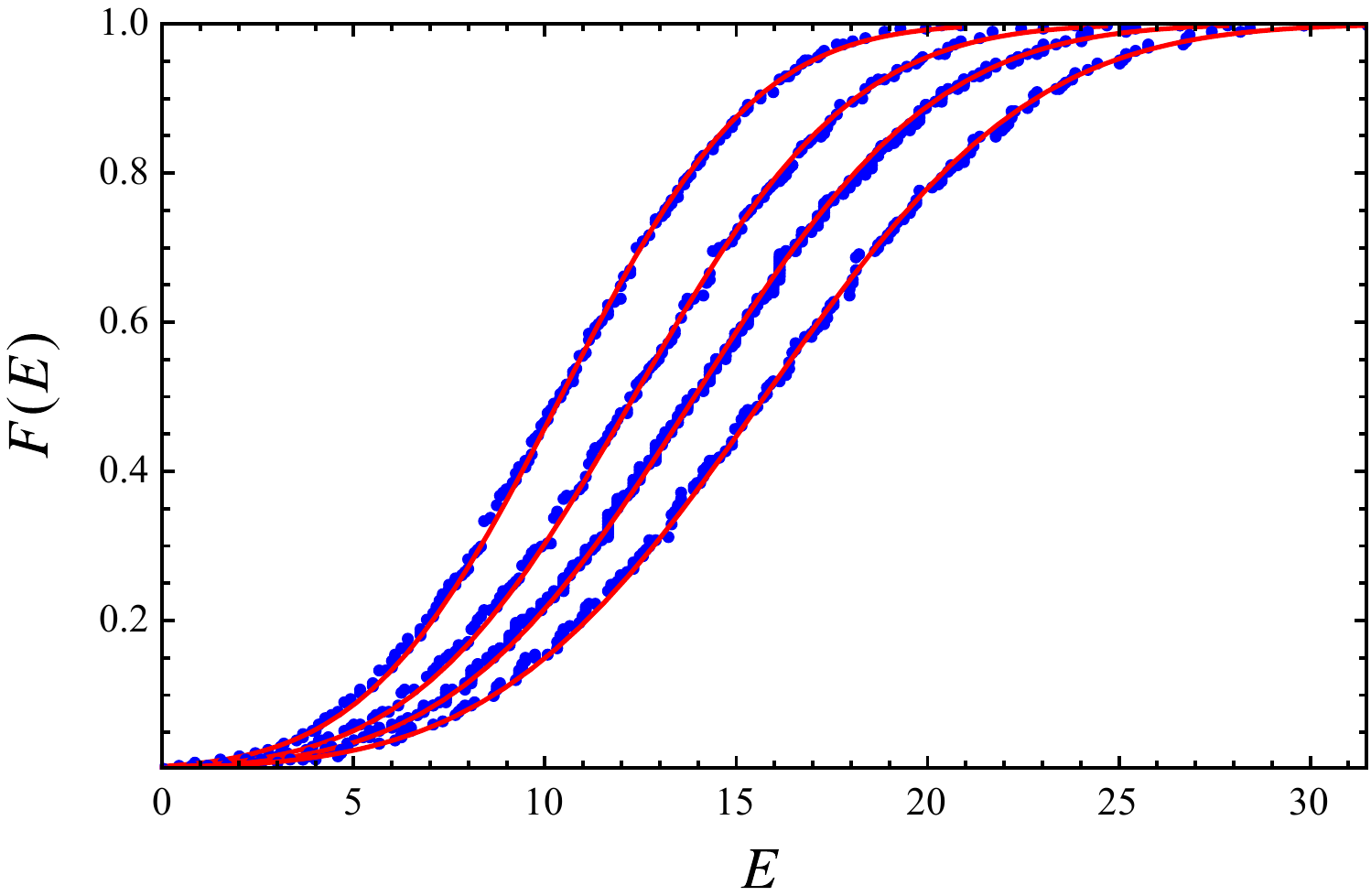}
  \caption{Blue dots: cumulative level density~\eqref{CLD} of the $\su(1|1)$ elliptic chain for
    $N=10$ and $\al=2,5,10,50$ (left to right). Continuous red lines: corresponding cumulative
    Gaussian distributions~\eqref{CGD} with parameters $\mu(\al)$ and $\si(\al)$ taken from the
    spectrum.}
  \label{fig.CLD}
\end{figure}
This proves that the level density of the elliptic $\su(1|1)$ chain becomes asymptotically
Gaussian as the number of spins tends to infinity. In particular, this holds in the limiting cases
$\al=0$ and $\al=\infty$, i.e., for the critical $X\!X$ model~\eqref{XX} and the $\su(1|1)$
Haldane--Shastry chain~\eqref{HS}. In order to better illustrate this property, it is convenient
to consider the (normalized) cumulative level density
\begin{equation}\label{CLD}
F(E)=2^{-N}\sum_{i;E_i\le E}d_i\,,
\end{equation}
where $E_1<\cdots<E_n$ are the distinct eigenvalues and $d_1,\dots,d_n$ their respective
degeneracies. In Fig.~\ref{fig.CLD} we have compared $F(E)$ for $N=10$ and several values of $\al$
with the corresponding cumulative Gaussian distribution
\begin{equation}\label{CGD}
G(E)=\frac12\,\bigg[1+\erf\bigg(\frac{E-\mu(\al)}{\sqrt 2\,\si(\al)}\bigg)\bigg]\,,
\end{equation}
where $\mu(\al)$ and $\si(\al)$ respectively denote the mean and the standard deviation of the
spectrum. It is apparent that, in spite of the relatively small value of $N$, the agreement
between the exact and the approximate cumulative distributions is remarkable (in fact, for
$N\gtrsim15$ both distributions are virtually indistinguishable).

In view of the previous discussion, it is of interest to compute in closed form the mean and the
standard deviation of the spectrum of the $\su(1|1)$ elliptic chain~\eqref{FG}-\eqref{newh}. In
the first place, the mean energy $\mu$ of any translation-invariant chain of the form~\eqref{FG}
is easily calculated from Eq.~\eqref{FGf}, taking into account that
\[
\tr(a^\dagger_ia_j)=2^{N-1}\de_{ij}\,.
\]
Indeed,
\[
\mu=\langle H\rangle=2^{-N}\tr H=-\frac N2\,h(0)=\frac N2\,\sum_{l=1}^{N-1}h(l)\,.
\]
From Eq.~\eqref{wpNsum}, which can be equivalently rewritten as
\begin{equation}
  \label{sumwp}
  \sum_{j=1}^{N-1}\wp_N(j)=\frac2{\al^2}\big(N\hat\eta_1-\eta_1)\,,
\end{equation}
and Eq.~\eqref{newh} we immediately obtain the following explicit formula for the mean energy of
the $\su(1|1)$ elliptic chain~\eqref{FG}-\eqref{newh}:
\begin{equation}
  \label{mu}
  \mu=\frac N{\pi^2}\sinh^2(\pi/\al)(\hat\eta_1-\eta_1)\,.
\end{equation}
The standard deviation of the spectrum of a translation-invariant chain~\eqref{FG}-\eqref{trinv}
can be computed in much the same way. Indeed, note to begin with that the trace of the product
$a_i^\dagger a_ja^\dagger_ka_l$ vanishes for all $i,j,k,l$ except in the following three
cases:\goodbreak
\begin{eqnarray*}\fl
  \tr(a_i^\dagger a_ia^\dagger_ia_i)&=\tr(a_i^\dagger a_i)=2^{N-1}\\\fl
  \tr(a_i^\dagger a_ia^\dagger_ja_j)&=2^{N-2}\quad (i\ne j)\\\fl
  \tr(a_i^\dagger a_ja^\dagger_ja_i)&=\tr(a_i^\dagger a_ia_ja^\dagger_j)
   =\tr(a_i^\dagger a_i)-\tr(a_i^\dagger a_ia^\dagger_ja_j)=2^{N-1}-2^{N-2}=2^{N-2}\quad(i\ne j)\,.
\end{eqnarray*}
Hence
\begin{eqnarray*}
  \big\langle H^2\big\rangle&=2^{-N}\tr(H^2)=\sum_{i,j,k,l=1}^Nh(|i-j|)h(|k-l|)\tr(a_i^\dagger
                              a_ja^\dagger_ka_l)\\
                            &=\frac N2\,h(0)^2+\frac14\,N(N-1)h(0)^2+\frac14\sum_{1\le i\ne j\le N}h(|i-j|)^2\\
                            &=\frac14\,N(N+1)\left(\sum_{j=1}^{N-1}h(j)\right)^2
                              +\frac N4\sum_{j=1}^{N-1}h(j)^2\,,
\end{eqnarray*}
where we have used Eq.~\eqref{sum21} with $h^2$ in place of $h$. We thus obtain
\[
\si^2=\big\langle H^2\big\rangle-\mu^2=\frac N4\left[\left(\sum_{j=1}^{N-1}h(j)\right)^2
  +\sum_{j=1}^{N-1} h(j)^2\right].
\]
Using Eqs.~\eqref{newh} and~\eqref{sumwp}, after a long but straightforward computation we arrive
at the following formula for the variance of the energy of the elliptic
chain~\eqref{FG}-\eqref{newh}:
\[
\si^2=N\,\bigg(\frac\al\pi\bigg)^4\sinh^4(\pi/\al)
\left[\frac14\sum_{j=1}^{N-1}\wp_N^2(j)+\frac1{\al^4}\,(\eta_1^2-N\hat\eta_1^2)\right]\,.
\]
The sum in the latter equation was evaluated in Ref.~\cite{FG14}, with the result:
\begin{eqnarray*}
\sum_{j=1}^{N-1}\wp_N^2(j)&=\frac1{12}\,\bigg(N-\frac65\bigg)g_2\biggl(\frac
N2,\frac{\iu\al}2\biggr) +\frac1{60}\,g_2\biggl(\frac 12,\frac{\iu\al}2\biggr)\\&=
\frac1{12\al^4}\bigg[\bigg(N-\frac65\bigg)g_2\biggl(\frac
12,\frac{\iu N}{2\al}\biggr) +\frac1{5}\,g_2\biggl(\frac 12,\frac\iu{2\al}\biggr)\bigg]\,,
\end{eqnarray*}
where we have used the homogeneity property of the Weierstrass invariant $g_2$:
\[
g_2(\la\om_1,\la\om_3)=\frac1{\la^4}\,g_2(\om_1,\om_3)\,.
\]
Setting
\[
g_2\equiv g_2\biggl(\frac 12,\frac\iu{2\al}\biggr)\,,\qquad
\hat g_2\equiv g_2\biggl(\frac 12,\frac{\iu N}{2\al}\biggr)
\]
we finally obtain the following exact formula for the variance of the spectrum of the
chain~\eqref{FG}-\eqref{newh}:
\begin{equation}
  \label{si2ell}
  \si^2=\frac{N}{\pi^4}\,\sinh^4(\pi/\al)\bigg[\bigg(N-\frac65\bigg)\frac{\hat g_2}{48}
  +\frac{g_2}{240}+\eta_1^2-N\hat\eta^2_1\bigg]\,.
\end{equation}

To conclude this section, we shall compute the limiting values of $\mu/N$ and $\si^2/N$ when
$N\to\infty$, which respectively coincide with the thermodynamic energy per particle $u$ and its
derivative with respect to $\be$ at $\be=0$. The first of these limits is a straightforward
consequence of Eq.~\eqref{eta1series} with $\om_1=1/2$ and $q=\exp(-N\pi/\al)\to0$:
\[
\lim_{N\to\infty}\frac{\mu}N=\frac1{\pi^2}\sinh^2(\pi/\al)\bigg(\frac{\pi^2}6-\eta_1\bigg)\,.
\]
Equating this expression to the value of $u$ at $\be=0$ obtained from Eq.~\eqref{u} with $J=1$,
i.e.,
\[
u(0)=\frac12\int_0^1\cE(p)\,\diff p
\]
we immediately obtain Eq.~\eqref{cEint}. When taking into account the definition~\eqref{cE} of
$\cE$, the latter equation becomes the remarkable identity
\[
\int_0^1\Big[\wp(p)-\big(\ze(p)-2\eta_1 p\big)^2\Big]\,\diff p=\frac{2\pi^2}3\,.
\]
Similarly, from Eq.~\eqref{eta1series} and the series
\[
g_2(\om_1,\om_3)=\bigg(\frac{\pi}{2\om_1}\bigg)^4\left(\frac43
  +320\sum_{n=1}^\infty\frac{n^3q^{2n}}{1-q^{2n}}\right)\,,\qquad
q\equiv\exp(\iu\pi\om_3/\om_1)\,,
\]
(cf.~\cite{SS03}) we easily obtain
\[
\lim_{N\to\infty}\frac{\si^2}N=\frac1{\pi^4}\,\sinh^4(\pi/\al)\bigg(\frac{g_2}{240}
+\eta_1^2-\frac{\pi^4}{30}\bigg)\,.
\]

\section{The infinite chain}\label{sec.infchain}
We next consider the $\su(1|1)$ supersymmetric version of the infinite Inozemtsev
chain~\cite{In90}, which we define as
\multiparteqlabel{Hinfg}
\numparts
\begin{equation}\label{Hinf}
  H=\frac J2\sum_{-\infty\le j\ne k\le\infty}g(j-k)(1-\sS_{jk})\,,   
\end{equation}
with
\begin{equation}\label{g}
g(x)=\bigg(\frac\pi\al\bigg)^2\sinh^{-2}(\pi x/\al)\,.
\end{equation}
\endnumparts
This model is closely related to the $N\to\infty$ limit of the elliptic
chain~\eqref{FG}-\eqref{newh}. Indeed, if $h$ is defined by Eq.~\eqref{newh} and $x\ne0$ is fixed,
from Eqs.~\eqref{wpseries2} (with $\om_1=-\iu\al/2$, $\om_3=N/2$) and~\eqref{eta1series} (with
$\om_1=1/2$, $\om_3=\iu N/(2\al)$) we easily obtain
\[
\lim_{N\to\infty} h(x)=\bigg(\frac\al\pi\bigg)^2\sinh^2(\pi/\al)\,g(x)\,.
\]
Thus, it should be expected that the infinite chain~\eqref{Hinf} is related to the thermodynamic
limit of the finite elliptic chain~\eqref{FG}-\eqref{newh}. In order to substantiate this
heuristic observation, consider more generally a translation-invariant chain of the
form~\eqref{Hinf}, defined by an arbitrary (smooth) even function $g$. Using Eq.~\eqref{Sij} we
can express the Hamiltonian of this chain in terms of fermion creation/annihilation operators $a_n$,
$a^\dagger_n$ as
\begin{equation}\label{Hinfa}
H=-J\sum_{j,k=-\infty}^\infty g(j-k)a^\dagger_ja_k\,,
\end{equation}
where
\[
g(0)\equiv-\sum_{0\ne l\in\ZZ}^\infty g(l)\,.
\]
We next introduce the Fourier-transformed operators
\[
c(p)=\sum_{n=-\infty}^\infty\e^{-2\pi\iu np}a_n\,,
\]
where $p\in(0,1)$ is again the physical momentum\footnote{Since the chain is invariant under
  integer translations, the momentum can be taken to belong to the ``Brillouin interval''
  $[0,2\pi)$.} in units of $2\pi$. From the identity
\[
\sum_{n=-\infty}^\infty\e^{2\pi\iu nx}=\de(x)
\]
it easily follows that the operators $c(p)$ satisfy the canonical anticommutation relations
\[
\{c(p),c(q)\}=\{c^\dagger(p),c^\dagger(q)\}=0,\qquad \{c^\dagger(p),c(q)\}=\de(p-q)\,.
\]
Applying the inverse Fourier transform
\[
a_n=\int_0^1\e^{2\pi\iu np}c(p)\,\diff p
\]
to the Hamiltonian~\eqref{Hinfg} after a straightforward calculation we obtain
\[
H=J\int_0^1\vep(p)\,c^\dagger(p) c(p)\,\diff p\,,
\]
where the dispersion relation $\vep(p)$ is given by
\begin{equation}\label{epsinf}
\vep(p)=\sum_{0\ne j\in\ZZ}\big(1-\e^{\pm2\pi\iu jp}\big)g(j)=2\sum_{j=1}^\infty\big[1-\cos(2\pi j
  p)\big]\,g(j)\,.
\end{equation}
In particular, $H$ is completely integrable, since the operators $c^\dagger(p)c(p)$ are a
commuting family of first integrals depending on the continuous parameter $p\in(0,1)$.
Alternatively, from the definition of $c(p)$ it easily follows that the operators
\[
\sum_{n=-\infty}^\infty a^\dagger_{n+k}a_n\,,\quad k\in\ZZ\,,
\]
form a denumerable family of commuting first integrals.

When $g$ is the function in Eq.~\eqref{g}, the latter sum can be evaluated using similar
techniques as in the finite (elliptic) case. Indeed, consider first the Fourier series
\[
\hat g(p)\equiv\sum_{0\ne l\in\ZZ}\e^{-2\pi\iu lp}\,g(l)\,,\qquad 0<p<1\,.
\]
In order to compute $\hat g$, we introduce the auxiliary function
\begin{equation}\label{Gzdef}
  G(z)=\sum_{j=-\infty}^\infty\e^{-2\pi\iu jp}\,g(z+j)
\end{equation}
dependent on the parameter $p\in(0,1)$, which clearly satisfies
\[
G(z+1)=\e^{2\pi\iu p}G(z)\,,\qquad G(z+\iu\al)=G(z)
\]
with $\exp(2\pi\iu p)\ne1$. Each of the terms in Eq.~\eqref{Gzdef} is analytic at the origin
except the one with $j=0$, whose Laurent series about the origin is $g(z)=z^{-2}+\Or(1)$. Thus
$G(z)=z^{-2}+\Or(1)$, and we can therefore apply Eqs.~\eqref{fz} and~\eqref{z0term} with $f=G$ and
$2\om_3=\iu\al$. The Fourier series $\hat g(p)$ is now evaluated by computing the constant term in
the Laurent series about the origin of $G(z)$ both directly and using Eq.~\eqref{z0term}. Indeed,
since
\[
g(z)=\frac1{z^2}-\frac{\pi^2}{3\al^2}+\Or(z^2)
\]
and the terms with $j\ne0$ in Eq.~\eqref{Gzdef} are regular at the origin, the Laurent series of
$G$ about the origin is given by
\[
G(z)=\frac1{z^2}-\frac{\pi^2}{3\al^2}+\sum_{0\ne j\in\ZZ}\e^{-2\pi\iu jp}\,g(j)+\Or(z)\equiv
\frac1{z^2}+\hat g(p)-\frac{\pi^2}{3\al^2}+\Or(z)\,.
\]
Comparing with Eq.~\eqref{z0term} we immediately obtain
\begin{equation}
  \label{FTfinal}
  \hat g(p)=\frac12\,\wp_1(\iu\al p)-\frac12\,
  \Big(\ze_1(\iu\al p)-2\eta_3p\Big)^2+\frac{\pi^2}{3\al^2}\,,\qquad 0<p<1\,.
\end{equation}
The sum $\ds\sum\limits_{0\ne l\in\ZZ}g(l)$ can now be computed by noting that, since the Fourier
series defining $\hat g$ is uniformly convergent for all $p$ by Weierstrass's test, $\hat g$ is a
continuous function of $p$. In particular
\begin{equation}\label{sinhsum}
\sum\limits_{0\ne l\in\ZZ}g(l)=\lim_{p\to0}\hat g(p)=\frac{2\eta_3}{\iu\al}+\frac{\pi^2}{3\al^2}\,,
\end{equation}
where the limit has been computed using Eqs.~\eqref{wpL} and~\eqref{zesigma}. Substituting
Eqs.~\eqref{FTfinal} and~\eqref{sinhsum} into Eq.~\eqref{epsinf} we finally obtain
\begin{equation}\label{vepim}
\vep(p)=
  \frac12\Big(\ze_1(\iu\al p)-2\eta_3p\Big)^2-\frac12\,\wp_1(\iu\al p)+\frac{2\eta_3}{\iu\al}\,,
\end{equation}
or, taking into account the homogeneity property~\eqref{homog} of the Weierstrass functions,
\[
\vep(p)=\frac1{2\al^2}\,\big[\wp(p)-(\ze(p)-2\eta_1p)^2-4\eta_1\big]
\equiv\bigg(\frac\pi\al\bigg)^2\sinh^{-2}\biggl(\frac\pi\al\biggr)\,\cE(p)\,,
\]
where $\cE(p)$ is the dispersion relation of the (finite) elliptic chain (cf.~Eq.~\eqref{cE}).
Thus the infinite hyperbolic chain~\eqref{Hinf} turns out to be equivalent to the thermodynamic
limit of the elliptic $\su(1|1)$ chain~\eqref{FG}-\eqref{newh} (up to a constant factor).
\begin{rem}
  The dispersion relation~\eqref{vepim} of the $\su(1|1)$ infinite chain~\eqref{Hinfg} coincides
  with the expression for the energy of the $1$-magnons with momentum $2\pi p$ ($\bmod\en2\pi$) of
  its $\su(2)$ counterpart which can be found in Ref.~\cite{In03}. This is no coincidence since,
  as in the finite case, it can be shown that the same is true for any translation-invariant chain
  of the form~\eqref{Hinf}.
\end{rem}
\begin{rem}
  The sum~\eqref{sinhsum} is of interest in the context of the AdS/CFT conjecture. Indeed, in
  Ref.~\cite{SS04} it is shown that the perturbative expansion of the (planar, two complex scalar
  fields) dilation operator of $\cN=4$ super Yang--Mills theory up to three loops can be obtained
  from the spectrum of the $\su(2)$ analog of the infinite hyperbolic chain~\eqref{Hinf}
  \[
  H=\frac18\sum_{-\infty\le j\ne l\le\infty}\sinh^{-2}\bigl(\ka(j-l)\bigr)(1-S_{jl})
  \]
  provided that the Yang--Mills coupling constant $\la$ is related to the parameter $\ka$ by
  \[
  \la = 4\pi^2\,\sum_{n=1}^\infty\sinh^{-2}(n\ka)
  \] 
  (cf.~\cite{SS04}, Eq.~(2.10)). Using Eq.~\eqref{sinhsum} with $\al=\pi/\ka$ we easily arrive at
  the following explicit formula for the coupling constant $\la$:
  \[
  \la=
  \frac{4\pi^2}{\ka^2}\,\left[\frac{\ka}{\iu\pi}\,\eta_3\biggl(\frac12,\frac{\iu\pi}{2\ka}\biggr)
    +\frac{\ka^2}6\right] =\frac{2\pi^2}3-4\eta_1\biggl(\frac12,\frac{\iu\ka}{2\pi}\biggr)\,.
   \]
   From Eqs.~\eqref{eta1series}-\eqref{2series} with $\om_1=1/2$, $\om_3=\iu\ka/(2\pi)$ we
   immediately obtain the expansion
   \[
   \la = 16\pi^2\sum_{n=1}^\infty\frac{n\,
     \e^{-2n\ka}}{1-\e^{-2n\ka}}=16\pi^2\sum_{n=1}^\infty\si_1(n)\e^{-2n\ka}\,,
   \]
   where $\si_1(n)$ denotes the number-theoretic divisor function
   \[
   \si_1(n)\equiv\sum_{j\ \mathrm{divides}\ n}j\,.
   \]
\end{rem}

\section{Conclusions and outlook}\label{sec.concout}
We have introduced the $\su(1|1)$ supersymmetric version of Inozemtsev's elliptic (finite) and
hyperbolic (infinite) spin chains, presented a proof of their integrability, and obtained their
exact solution. This is rather unexpected, in view of the fact that no rigorous proof of the
integrability or complete solution of the apparently simpler $\su(2)$ version of these models is
known. Taking advantage of the explicit knowledge of the spectrum, we have been able to compute in
closed form the thermodynamic functions, showing that at low temperatures the $\su(1|1)$ elliptic
chain is essentially equivalent to a critical $X\!X$ model. We have also rigorously proved that
the spectrum is normally distributed in the thermodynamic limit, as is typically the case with
spin chains of Haldane--Shastry type. In fact, these results also apply to the standard $X\!X$
model at a critical value of the magnetic field, since this model is obtained from $\su(1|1)$
elliptic chain when the imaginary period of the interaction strength tends to zero.

Our results suggest several new developments and open problems. In the first place, it would be of
interest to determine whether the elliptic $\su(1|1)$ and $\su(2)$ chains are equivalent in the
thermodynamic limit, as we have shown to be the case for their Haldane--Shastry counterparts.
Another natural problem is the extension of the present work to elliptic chains of $\su(m|n)$
type. Note, in this respect, that in the thermodynamic limit and at low temperature the $\su(m|1)$
Haldane--Shastry chain is equivalent to a model of $m$ species of non-interacting fermions, with
the same dispersion relation as its $\su(2)$ and $\su(1|1)$ versions~\cite{BBS08}. A third line of
future research opened up by our exact solution of the $\su(1|1)$ elliptic chain is the study of
properties of its spectrum of relevance in the characterization of integrability vs.~quantum chaos
(nearest-neighbor spacing distribution, spectral noise, etc.), as has been done for spin chains of
Haldane--Shastry type (see, e.g., Ref.~\cite{BB06,BFGR08epl,BFGR09power,BB09}). Finally, another
potential application of our results is the computation of the entanglement entropy of the ground
state of the elliptic chain~\eqref{FG}-\eqref{newh}, using the method outlined in Ref.~\cite{LR09}
for the $X\!X$ model in a constant magnetic field.

\ack This work was supported in part by Spain's MINECO under grant no.~FIS2011-22566.

\appendix
\section{Evaluation of some limits involving the Weierstrass elliptic function}
\label{app.limits}
Our starting point shall be the two trigonometric series
\begin{eqnarray}\fl
  \label{wpseries}
  \wp(z;\om_1,\om_3)=-\frac{\eta_1(\om_1,\om_3)}{\om_1}+\frac{\pi^2}{4\om_1^2}\,
  \sin^{-2}\Bigl(\frac{\pi z}{2\om_1}\Bigr)-\frac{2\pi^2}{\om_1^2}\sum_{n=1}^\infty
  \frac{nq^{2n}}{1-q^{2n}}\,\cos\Bigl(\frac{n\pi z}{\om_1}\Bigr)\\\fl
  \eta_1(\om_1,\om_3)=\frac{\pi^2}{12\om_1}-\frac{2\pi^2}{\om_1}\sum_{n=1}^\infty
  \frac{q^{2n}}{(1-q^{2n})^2}\,,
  \label{eta1series}
\end{eqnarray}
where
\[
q\equiv\e^{\iu\pi\om_3/\om_1}\,,
\]
$|\Im(z/\om_1)|<2\Im(\om_3/\om_1)$, and $z\ne m\om_1+n\om_3$ for all $m,n\in\ZZ$
(see~\cite{OLBC10}, Eqs.~(23.8.1)-(23-8.5)). Both series can be combined by noting that
\begin{equation}\label{2series}\fl
\sum_{n=1}^\infty \frac{q^{2n}}{(1-q^{2n})^2}=\sum_{n=1}^\infty\sum_{m=1}^\infty
m\,q^{2mn}=\sum_{m,n=1}^\infty n\,q^{2mn}=\sum_{n=1}^\infty n\sum_{m=1}^\infty q^{2mn}
=\sum_{n=1}^\infty\frac{nq^{2n}}{1-q^{2n}}\,,
\end{equation}
where all the rearrangements are justified by the absolute convergence of the double series (note
that $|q|<1$ on account of the condition $\Im(\om_3/\om_1)>0$). Inserting~\eqref{eta1series}
into~\eqref{wpseries} we thus obtain
\begin{equation}\fl
  \label{wpseries2}
  \wp(z;\om_1,\om_3)=\frac{\pi^2}{\om_1^2}\left[-\frac{1}{12}+\frac{1}{4}\,
  \sin^{-2}\Bigl(\frac{\pi z}{2\om_1}\Bigr)+4\sum_{n=1}^\infty
  \frac{nq^{2n}}{1-q^{2n}}\,\sin^2\Bigl(\frac{n\pi z}{2\om_1}\Bigr)\right]\,.
\end{equation}
Equation~\eqref{toin} immediately follows by taking $\om_1=N/2$, $\om_3=\iu\al/2$ and letting
$\al\to\infty$, i.e., $q=\exp(-\pi\al/N)\to0$. In order to examine the behavior of $\wp_N(x)$ as
$\al\to0$, we first note that
\[
\wp_N(x)\equiv\wp\biggl(x;\frac N2,\frac{\iu\al}2\biggr)=\wp\biggl(x;-\frac{\iu\al}2,\frac
N2\biggr)\,.
\]
If $1\le x\le N-1$, from Eq.~\eqref{wpseries2} with $\om_1=-\iu\al/2$, $\om_3=N/2$, and thus
$q=\exp(-N\pi/\al)\to0$, we have
\begin{eqnarray}\fl
  \e^{2\pi/\al}\bigg(\frac{\al^2}{4\pi^2}\,\wp_N(x)
  -\frac1{12}\bigg)=&\frac{q^{2(x-1)/N}}{(1-q^{2x/N})^2}
  +\sum_{n=1}^\infty\frac{nq^{2[n(N-x)-1]/N}(1-q^{2nx/N})^2}{1-q^{2n}}\nonumber\\
  &\underset{q\to0}{\longrightarrow}\de_{1,x}+\de_{N-1,x}\,.
  \label{wpal0}
\end{eqnarray}

We shall next prove that the Hamiltonian~\eqref{FG}-\eqref{newh} smoothly interpolates between the
Heisenberg and the Haldane--Shastry Hamiltonians as the parameter $\al$ ranges from zero to
infinity. In other words, if $h(x)$ denotes the function defined by Eq.~\eqref{newh} we shall show
that
\[
\lim_{\al\to0}h(x)=\de_{1,x}+\de_{N-1,x}\,,\qquad
\lim_{\al\to\infty}h(x)=\frac{\pi^2}{N^2}\,\sin^{-2}\biggl(\frac{\pi x}N\biggr)\,,
\]
where in the former limit $1\le x\le N-1$. Consider, to begin with, the limit $\al\to0$.
From Eq.~\eqref{eta1series} with $\om_1=1/2$, $\om_3=\iu N/(2\al)$ and, therefore,
$q=\exp(-N\pi/\al)\to0$ we have
\[
\hat\eta_1\equiv\eta_1\biggl(\frac12,\frac{\iu N}{2\al}\biggr)=\frac{\pi^2}6+\Or\bigl(\e^{-2N\pi/\al}\bigr)\,.
\]
Thus
\begin{eqnarray*}
  h(x)&=\bigg(\frac\al\pi\bigg)^2\sinh^2(\pi/\al)\bigg(\wp_N(x)
  -\frac{\pi^2}{3\al^2}\bigg)+\Or\bigl(\e^{-2\pi(N-1)/\al}\bigr)\\
  &=\big(1-\e^{-2\pi/\al}\big)^2\cdot\e^{2\pi/\al}\bigg(\frac{\al^2}{4\pi^2}\,\wp_N(x)
  -\frac1{12}\bigg)+\Or\bigl(\e^{-2\pi(N-1)/\al}\bigr)\,,
\end{eqnarray*}
which indeed tends to $\de_{1,x}+\de_{N-1,x}$ for $1\le x\le N-1$ as $\al\to0$ on account of
Eq.~\eqref{wpal0}.

Consider now the $\al\to\infty$ limit. From the homogeneity property of the Weierstrass zeta
function (see Eq.~\eqref{homog}) we have
\[
\hat\eta_1=\frac{\iu\al}{N}\,\eta_3\biggl(\frac12,\frac{\iu\al}{2N}\biggr)
=\frac{\al}{N}\,\bigg[\pi-\frac\al N\,\eta_1\biggl(\frac12,\frac{\iu\al}{2N}\biggr)\bigg]\,,
\]
where in the last step we have used Legendre's relation (cf.~\cite{OLBC10}, Eq.~(23.2.14))
  \begin{equation}\label{Legendre}
\om_3\eta_1(\om_1,\om_3)-\om_1\eta_3(\om_1,\om_3)=\frac{\iu\pi}2\,.
  \end{equation}
From the series~\eqref{eta1series} we easily obtain
\[
\lim_{\al\to\infty}\eta_1\biggl(\frac12,\frac{\iu\al}{2N}\biggr)=\frac{\pi^2}6\,,
\]
and therefore
\[\fl
h(x)=\bigg(\frac\al\pi\bigg)^2\sinh^2(\pi/\al)\bigg(\wp_N(x)
+\frac2{N^2}\,\eta_1\biggl(\frac12,\frac{\iu\al}{2N}\biggr)-\frac{2\pi}{N\al}\bigg)
\underset{\al\to\infty}{\longrightarrow}\frac{\pi^2}{N^2}\sin^{-2}\biggl(\frac{\pi x}N\biggr)
\]
on account of Eq.~\eqref{toin}.

We shall finally prove that the dispersion relation~\eqref{cE} of the elliptic $\su(1|1)$
chain~\eqref{FG}-\eqref{newh} does tend to the dispersion relations of the critical~$X\!X$
model~\eqref{XX} and the $\su(1|1)$ Haldane--Shastry chain~\eqref{HS} respectively as $\al\to0$
and $\al\to\infty$. In other words, we shall show that
\[
\lim_{\al\to0}\cE(p)=4\sin^2(\pi p)\,,\qquad \lim_{\al\to\infty}\cE(p)=2\pi^2p(1-p)
\]
(cf.~Eqs.~\eqref{dispHS}-\eqref{dispXX}). Consider, to begin with, the former of these limits.
Using Eqs.~\eqref{eta1series}-\eqref{wpseries2} with $\om_1=1/2$, $\om_3=\iu\al/2$ and, hence,
$q=\exp(-\pi/\al)\to0$ we readily obtain
\[
\wp(p)-4\eta_1=\pi^2\cot^2(\pi p)+16\pi^2\e^{-2\pi/\al}\big(1+\sin^2(\pi
p)\big)+\Or\bigl(\e^{-4\pi/\al}\bigr)\,.
\]
Similarly, from the series
\begin{equation}\label{zeseries}
\fl
\ze(z;\om_1,\om_3)=\frac{\eta_1(\om_1,\om_3)}{\om_1}\,z+\frac{\pi}{2\om_1}\cot\biggl(\frac{\pi
  z}{2\om_1}\biggr)+\frac{2\pi}{\om_1}\sum_{n=1}^\infty\frac{q^{2n}}{1-q^{2n}}\sin\biggl(\frac{n\pi
  z}{\om_1}\biggr)\,,
\end{equation}
valid for $|\Im(z/\om_1)|<2\Im(\om_3/\om_1)$ and $z\ne m\om_1+n\om_3$ for all $m,n\in\ZZ$
(cf~\cite{OLBC10}, Eq.~(23.8.2)) we have
\[
\ze(p)-2\eta_1 p=\pi\cot(\pi p)+4\pi\e^{-2\pi/\al}\sin(2\pi p)+\Or\bigl(\e^{-4\pi/\al}\bigr)\,.
\]
Hence
\[
  \wp(p)-(\ze(p)-2\eta_1 p)^2-4\eta_1
  =32\pi^2\e^{-2\pi/\al}\sin^2(\pi p)+\Or\bigl(\e^{-4\pi/\al}\bigr)\,,
\]
and multiplying by the factor $\sinh^2(\pi/\al)/(2\pi^2)$ we easily obtain
\begin{equation}
  \label{cEp0}
  \cE(p)=4\sin^2(\pi p)+\Or\bigl(\e^{-2\pi/\al}\bigr)
  \underset{\al\to0}{\longrightarrow}4\sin^2(\pi p)\,.
\end{equation}
Consider, finally, the $\al\to\infty$ limit. Using the homogeneity relations~\eqref{homog} we can
write
\begin{equation}\label{cEi}\fl
\cE(p)=\frac12\bigg(\frac\al\pi\bigg)^2\sinh^2(\pi/\al)\bigg[\Big(\ze_1(\iu\al
p)-2\eta_3\bigl(\tfrac12,\tfrac{\iu\al}2\bigr)p\Big)^2-\wp_1(\iu\al
p)+\frac4{\iu\al}\,\eta_3\biggl(\frac12,\frac{\iu\al}2\biggr)\bigg]\,,
\end{equation}
where $\wp_1$ and $\ze_1$ denote the Weierstrass functions with periods $1$ and $\iu\al$. From
Eqs.~\eqref{eta1series}, \eqref{wpseries2} and~\eqref{zeseries} with $\om_1=1/2$,
$\om_3=\iu\al/2$, so that $q=\exp(-\pi\al)\to0$, and the Legendre relation~\eqref{Legendre}, we
now obtain
\begin{eqnarray*}
  \wp_1(\iu\al p)=-\frac{\pi^2}3+\Or\bigl(\e^{-2\pi\al\min(p,1-p)}\bigr)\,,\\
  \ze_1(\iu\al p)-2\eta_3\bigl(\tfrac12,\tfrac{\iu\al}2\bigr)p=2\pi\iu\,p-\iu\pi\coth(\pi\al p)
  +\Or\bigl(\e^{-2\pi \al(1-p)}\bigr)\\
  \hphantom{\ze_1(\iu\al p)-2\eta_3\bigl(\tfrac12,\tfrac{\iu\al}2\bigr)p}
  =\pi\iu(2p-1)+\Or\bigl(\e^{-2\pi\al\min(p,1-p)}\bigr)\\
  \frac4{\iu\al}\,\eta_3\biggl(\frac12,\frac{\iu\al}2\biggr)=
  4\eta_1\biggl(\frac12,\frac{\iu\al}2\biggr)-\frac{4\pi}\al=
  \frac{2\pi^2}3+\Or(\al^{-1})\,.
\end{eqnarray*}
These asymptotic relations and Eq.~\eqref{cEi} immediately lead to
\begin{equation}\label{cEpinf}\fl
\cE(p)=\frac{\pi^2}2\big[1-(2p-1)^2\big] +\Or\bigl(\al^{-1}\bigr)=2\pi^2p(1-p)
+\Or\bigl(\al^{-1}\bigr)\underset{\al\to\infty}{\longrightarrow}2\pi^2p(1-p)\,,
\end{equation}
as claimed. Note that the convergence of $\cE(p)$ to the Haldane--Shastry dispersion
relation~\eqref{dispHS} as $\al\to\infty$ is much slower than its convergence to the $X\!X$
dispersion relation~\eqref{dispXX} as $\al\to0$, as can be seen from
Eqs.~\eqref{cEp0}-\eqref{cEpinf} (and is also apparent from Fig.~\ref{fig.cEp}).
\section{Quasi-periodic functions}\label{app.qpf}
In this section we derive several key results on quasi-periodic functions needed for the explicit
evaluation of the dispersion relation of the elliptic chain~\eqref{FG}-\eqref{newh} and its
infinite (hyperbolic) counterpart~\eqref{Hinf}-\eqref{g}.

By definition, a function $f(z)$ is (strictly) quasi-periodic with half-periods $\om_1$ and
$\om_3$ provided that
\begin{equation}\label{fqp}
f(z+2\om_1)=\e^{2\pi\iu p}f(z)\,,\qquad f(z+2\om_3)=f(z)\,,
\end{equation}
where $p\notin\ZZ$ is a real constant. Since the periods are assumed to be independent, i.e.,
$\Im(\om_3/\om_1)\ne0$, from now on we shall suppose without loss of generality that
$\Im(\om_3/\om_1)>0$. One of the most basic results about quasi-periodic functions is the
following immediate consequence of Liouville's theorem in analytic function theory: \emph{if a
  quasi-periodic function $f$ is analytic in the closed period parallelogram}
  \[
  \pi_{\om_1,\om_3}=\big\{2x\om_1+2y\om_3\mid0\le x,y\le 1\big\}\,,
  \]
  \emph{then it is identically zero.} Indeed, by the analiticity hypothesis $f$ is bounded in the
  compact set $\pi_{\om_1,\om_3}$. The quasi-periodicity conditions then imply that $f$ is entire
  and bounded, since for every $z\in\CC$ we can always find two integers $m,n$ such that
  $z-2m\om_1-2n\om_3=z_0\in\pi_{\om_1,\om_3}$, and hence
  \[
  |f(z)|=|\e^{2\pi\iu mp}f(z_0)|=|f(z_0)|\,.
  \]
  By Liouville's theorem, $f$ reduces to a constant $c$, which must vanish on account of the
  conditions $c=\e^{2\pi\iu p}c$ and $p\not\in\ZZ$.

  The previous result implies that, just as elliptic (doubly periodic) functions, quasi-periodic
  functions are essentially determined by their singularities. We shall only need here a very
  simple application of this idea, which we shall discuss next. More precisely, suppose that $f$
  is a quasi-periodic function whose only singularities are double poles on the period lattice
  $2m\om_1+2n\om_3$ ($m,n\in\ZZ$), and whose Laurent series about the origin
  is
  \[
  f(z)=\frac1{z^2}+\Or(1)\,.
  \]
  Let us assume, for simplicity, that $2\om_1=1$ (otherwise, it suffices to replace $f(z)$ by
  $f(\om_1 z)$ and $\om_3$ by $\om_3/\om_1$ in what follows). We shall then show that
\begin{eqnarray}\fl
  f(z)=\frac{\si_1(\om_3p+z)}{\si_1(\om_3p-z)}\,\e^{-A(p)z}\bigg[&\wp_1(z)-\wp_1(\om_3 p)\nonumber\\\fl
  &+B(p)
    \bigg(\ze_1(z)-\ze_1(z+\om_3p)+\ze_1(2\om_3p)-\ze_1(\om_3p)\bigg)\bigg]\,,
    \label{fz}
\end{eqnarray}
with
\begin{equation}\label{ApBp}
  A(p)=2\eta_3(1/2,\om_3)p\equiv 2\eta_3 p\,,
  \qquad B(p)=2\Big(\eta_3 p-\ze_1(\om_3p)\Big)\,.
\end{equation}
Here
\[\fl
\wp_1(z)\equiv\wp(z;1/2,\om_3)\,,\qquad \ze_1(z)\equiv\ze(z;1/2,\om_3)\,,\qquad
\si_1(z)\equiv\si(z;1/2,\om_3)\,,
\]
where $\si(z;\om_1,\om_3)$ denotes the Weierstrass sigma function~{\upshape\cite{WW27}}.
Indeed, let us denote by $f_0$ the RHS of Eq.~\eqref{fz}. The term in square brackets in $f_0$ is
clearly periodic, with periods $1$ and $2\om_3$, due to the identities
  \[
  \ze(z+2\om_i;\om_1,\om_3)=\ze(z;\om_1,\om_3)+2\eta_i(\om_1,\om_3)\,.
  \]
  On the other hand, from the relation
  \[
  \si(z+2\om_i;\om_1,\om_3)=-\exp\big[2\eta_i(\om_1,\om_3)(z+\om_i)\big]\,\si(z;\om_1,\om_3)
  \]
  and Legendre's identity~\eqref{Legendre} it follows that the factor
  \[
  g(z)\equiv\frac{\si_1(\om_3p+z)}{\si_1(\om_3p-z)}\,\e^{-A(p)z}
  \]
  satisfies
  \[
  g(z+1)=\e^{2\pi\iu p}\,g(z)\,,\qquad g(z+2\om_3)=g(z)\,.
  \]
  Thus $f_0$ is quasi-periodic, with the same half-periods as $f$. Furthermore, $f_0$ is analytic
  on the whole complex plane except on the period lattice. Indeed, the simple zero of
  $\si_1(\om_3p-z)$ at $\om_3p$ (and congruent points) is canceled by the zero of the term in
  square brackets, while the simple pole of the latter term due to $\ze_1(z+\om_3p)$ at
  $z=-\om_3p$ (and points congruent to it) is canceled by the simple zero of $\si_1(z+\om_3p)$.
  The behavior at the origin of $f_0$ can be easily determined using the relation
  \begin{equation}\label{zesigma}
    \ze(z;\om_1,\om_3)=\frac{\si'(z;\om_1,\om_3)}{\si(z;\om_1,\om_3)}=\frac1{z}+\Or(z^3)\,,
  \end{equation}
  from which it follows that
  \begin{eqnarray*}
  \frac{\si_1(\om_3p+z)}{\si_1(\om_3p-z)}
  &=\frac{\si_1(\om_3p)+\si_1'(\om_3p)z+\Or(z^2)}{\si_1(\om_3p)-\si_1'(\om_3p)z+\Or(z^2)}=
  \frac{1+\ze_1(\om_3p)z+\Or(z^2)}{1-\ze_1(\om_3p)z+\Or(z^2)}\\
  &=1+2\ze_1(\om_3p)z+\Or(z^2)\,.
  \end{eqnarray*}
  Using this identity it is straightforward to derive the principal part of $f_0$ at the origin,
  namely
  \[
  f_0(z)=\frac1{z^2}+\frac{B(p)-A(p)+2\ze_1(\om_3p)}{z}+\Or(1)\,.
  \]
  Since the coefficient of $1/z$ in the latter series vanishes identically on account of the
  definitions of $A(p)$ and $B(p)$, both sides of~\eqref{fz} have the same principal part at the
  origin and hence, by quasi-periodicity, at all points congruent to it. Hence the difference
  $f-f_0$ is strictly quasi-periodic and entire, and therefore vanishes identically by the basic
  property of quasi-periodic functions proved at the beginning of this appendix.

  The last result needed for the evaluation of the dispersion relations of the elliptic and
  hyperbolic $\su(1|1)$ chains is the following explicit formula for the constant term in the
  Laurent expansion about the origin of the function $f(z)$ in Eq.~\eqref{fz}:
  \begin{equation}
    \label{z0term}
    \lim_{z\to0}\bigg(f(z)-\frac1{z^2}\bigg)=\frac12\,\wp_1(2\om_3p)
    -\frac12\Big(\ze_1(2\om_3p)-2\eta_3p\Big)^2\,.
\end{equation}
To prove this formula, note that
\[
\frac{\si_1(\om_3p+z)}{\si_1(\om_3p-z)}=1+2\ze_1(\om_3p)z+2\ze_1(\om_3p)^2z^2+\Or(z^3)\,,
\]
from which it easily follows that
\[
\e^{-A(p)z}\frac{\si_1(\om_3p+z)}{\si_1(\om_3p-z)}=1-B(p)z+\frac12\,B(p)^2z^2+\Or(z^3)\,.
\]
Using this expansion and the Laurent series about the origin of the term in square brackets in
Eq.~\eqref{fz}, namely
\[
\frac1{z^2}+\frac{B(p)}z+B(p)\big[\ze_1(2\om_3p)-2\ze_1(\om_3p)\big]-\wp_1(\om_3p)+\Or(z)\,,
\]
we readily obtain
\[
f(z)=\frac1{z^2}+B(p)\big[\ze_1(2\om_3p)-2\ze_1(\om_3p)\big]-\wp_1(\om_3p)-\frac12\,B(p)^2+\Or(z)\,.
\]
We thus have
\begin{eqnarray*}
\fl\lim_{z\to0}\bigg(f(z)-\frac1{z^2}\bigg)&=
  B(p)\Big(\ze_1(2\om_3p)-2\ze_1(\om_3p)\Big)-\wp_1(\om_3p)-\frac12\,B(p)^2\\
\fl&=2\ze_1(2\om_3p)\Big(\eta_3p-\ze_1(\om_3p)\Big)+2\ze_1(\om_3p)^2-2\eta_3^2p^2-\wp_1(\om_3p)\,,
\end{eqnarray*}
where we have used the definition~\eqref{ApBp} of $B(p)$. Equation~\eqref{z0term} now follows
straightforwardly from the duplication formulas
\[
\wp(z)=-\frac12\,\wp(2z)+\frac18\,\left(\frac{\wp''(z)}{\wp'(z)}\right)^2\,,\qquad
\ze(z)=\frac12\,\ze(2z)-\frac14\,\frac{\wp''(z)}{\wp'(z)}\,,
\]
where $\wp(z)\equiv\wp(z;\om_1,\om_3)$ and similarly $\ze(z)$ (see, e.g.,~\cite{OLBC10}).

\section*{References}


\end{document}